%% file: main.tex
\documentclass[%
reprint,
superscriptaddress,
nofootinbib,
nolongbibliography,
 amsmath,amssymb,
 aps,
prd,
floatfix,
]{revtex4-2}

\usepackage{graphicx}
\usepackage{dcolumn}
\usepackage{bm}
\usepackage{xcolor}

\definecolor{maroon}{RGB}{192,0,0}
\usepackage[colorlinks=true, citecolor=maroon,
linkcolor=maroon,
urlcolor=blue]{hyperref}
\usepackage{times}
\usepackage{tikz}
\usepackage{multirow}
\usepackage{amsmath}	
\usepackage{amssymb}    
\usepackage{amssymb}	
\usepackage{hyperref}
\usepackage{contour}
\usepackage{ulem}
\usepackage{float}
\usepackage{pgfplots}
\pgfplotsset{compat=1.18}
\usepackage{enumitem}
\usepackage{comment}

\newcommand{\rockstar}{\textsc{Rockstar}}
\newcommand{\oasis}{\textsc{Oasis}}

\newcommand{\gadget}{\textsc{GADGET-2}}

\newcommand{\hmsun}{h^{-1}\ {\rm M_{\odot}}}

\newcommand{\hMpc}{h^{-1}\ {\rm Mpc}}

\newcommand{\hGpc}{h^{-1}\ {\rm Gpc}}

\newcommand{\Morb}{M_{\rm orb}}

\newcommand{\dsc}{\delta_{\rm sc}}

\newcommand{\hMsun}{h^{-1}\ M_\odot}

\newcommand{\Rball}{R_{\rm ball}}
\newcommand{\forb}{f_{\rm orb}}

\newcommand{\Omegam}{\Omega_{\rm m}}

\begin{document}


\title[OASIS]{OASIS: A Fast Dynamical Halo Finding Algorithm}


\author{Bayli Hayes}
    \email{baylihayes@arizona.edu}
    \affiliation{Department of Physics, University of Arizona, Tucson, AZ 85721, USA}
\author{Edgar M. Salazar}
    \affiliation{Department of Physics, University of Arizona, Tucson, AZ 85721, USA}
\author{Eduardo Rozo}
    \email{erozo@arizona.edu}
    \affiliation{Department of Physics, University of Arizona, Tucson, AZ 85721, USA}
\author{Benedikt Diemer}
    \affiliation{Department of Astronomy, University of Maryland, College Park, MD 20742, USA}
\author{Calvin Osinga}
    \affiliation{Department of Astronomy, University of Maryland, College Park, MD 20742, USA}   
    
\author{John Hugon}
    \affiliation{Department of Physics, University of Arizona, Tucson, AZ 85721, USA}
\author{Vadim Bernshteyn}
    \affiliation{Department of Physics, University of Arizona, Tucson, AZ 85721, USA}
\author{Ze'ev Vladimir}
    \affiliation{Department of Astronomy, University of Maryland, College Park, MD 20742, USA}
    \affiliation{Department of Astronomy and Steward Observatory, University of Arizona, 933 North Cherry Avenue, Tucson, AZ 85721, USA }

\date{\today}

\begin{abstract}
    In recent work (\citet{GarciaSalazar2022}), we advocated for defining a \textit{dynamical halo} as the collection of particles that orbit their self-generated potential. Here we introduce the Orbiting-mass ASsIgnment Scheme (\oasis), a fast energy-based algorithm for constructing dynamical halo catalogs. Using single-snapshot information, \oasis\ achieves $\simeq 9\%$ scatter in orbiting mass estimates relative to the {\sc Sparta} particle-tracking algorithm. As an illustration of the applications enabled by \oasis, we use the Quijote and Erebos simulation suites to show that the \mbox{$z=0$} dynamical halo mass function is nearly universal with respect to cosmology, and that it is well described by a modified Press--Schechter form with an effective collapse threshold $\dsc = 1.54$ at intermediate masses. However, this description fails at high masses.
\end{abstract}

\maketitle


\input{sections/intro}

\input{sections/data}
\input{sections/algorithm}
\input{sections/validation}


\input{sections/massfunction}

\input{sections/summary}

\section*{Data Availability}
The data underlying this article will be shared on reasonable request to the corresponding author. The code used to generate the halo catalogs in this paper is available at \url{https://github.com/baylihayes/oasis} (v1.3.0).

\vspace{\baselineskip}
\begin{acknowledgments}
E.S. and E.R. were supported by NSF grant 2206688.  B.H. is funded by NSF grant 2605463. E.R. is also funded by DOE grant DE-SC0009913. This research was supported in part by the National Science Foundation under Grant numbers 2206690 and 2338388.
\end{acknowledgments}


\bibliography{database}

\end{document}

%% file: sections/intro.tex
\section{Introduction}
\label{sec:intro}

Dark matter halos have long been considered the fundamental building blocks of large-scale structure \cite{Cooray-Sheth}. Despite their importance, there is no unique definition of halos in the literature.  The most popular is the spherical overdensity criterion: a halo is a sphere whose mean density is some factor $\Delta$ of the mean matter density of the Universe.  The overdensity factor is often taken to be the virial overdensity estimated using spherical collapse \citep{BryanNorman1998}.  However, there are no features in the halo density profiles at these radii, bringing into question the definition's underlying physical motivation. 

Over the past decade, multiple authors have suggested moving the boundary of a halo outwards to better coincide with features in the density or velocity profile  of halos \citep[e.g.][]{GarciaRozo2019,fong_21,Tomooka2020}. The first of these is  the \it splashback radius \rm \citep{diemerkravtsov14,moreetal15,mansfield_etal17},
i.e. the apocentric distance that particles reach after experiencing their first pericentric passage.  This radius manifests as a steepening of the slope of matter density as a function of halo radius \citep{adhikari2014}, and its use led to significant insights on halo structure and galaxy evolution \cite{Diemer2020, Diemer2021,mansfield_kravtsov_20}.  Nevertheless, splashback-based halo definitions still suffer from a significant drawback: much of the mass interior to this boundary is infalling towards the halo for the first time, making it dynamically distinct from the material orbiting the halo. Moreover, defining halos using a spherical boundary ignores the triaxial structure of the halo.

\begin{figure*}
    \centering
    \includegraphics[width=0.95\linewidth]{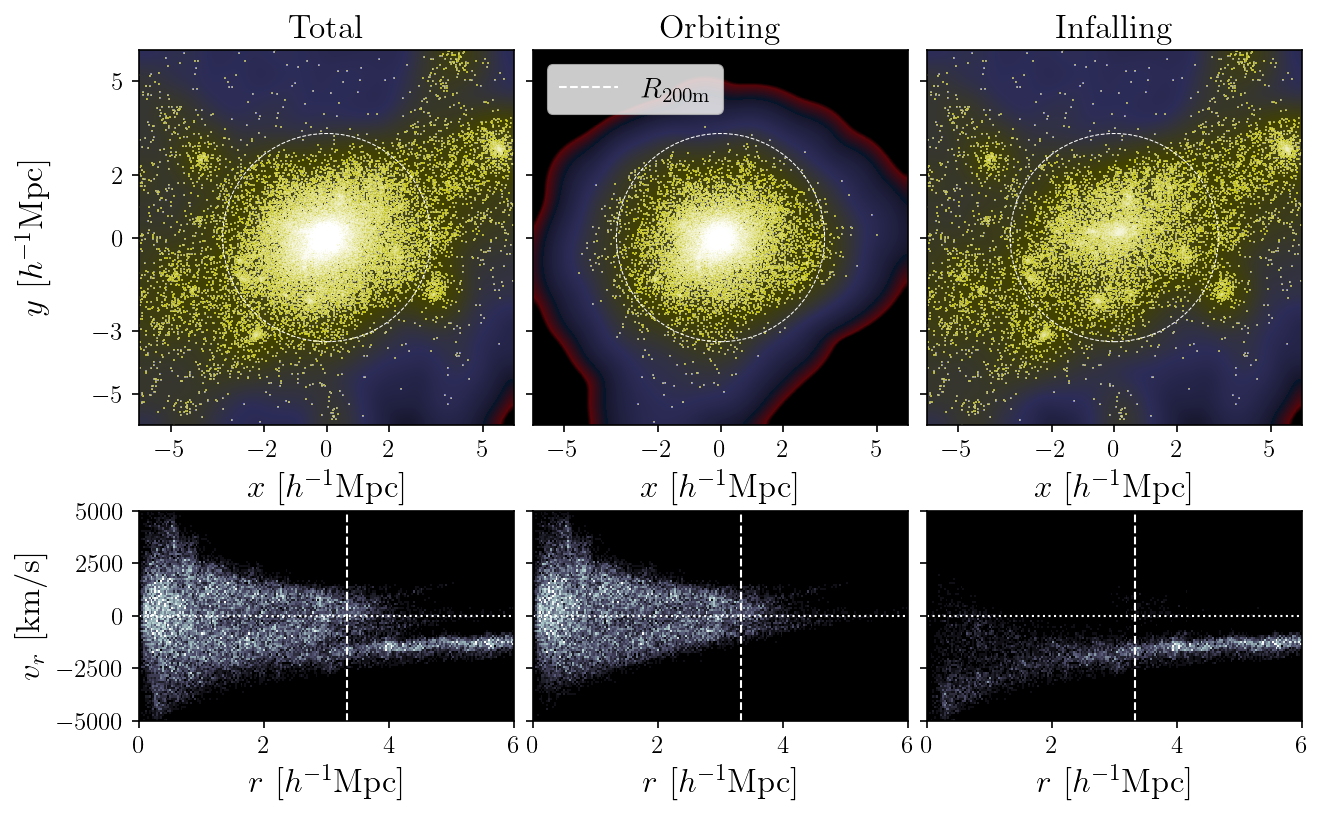}
    \caption{Dark matter particle distribution in the vicinity of the most massive halo in the Banerjee simulation (left), separated into orbiting (middle) and infalling (right) populations according to the accretion-time criterion from Paper I \cite{GarciaSalazar2022}. {\bf Top:} Configuration space $y-x$ projection. The dashed circle has a radius of $R_{\rm 200m}$ for this halo. The density distribution was smoothed with a wide Gaussian filter and overlaid in a logarithmic color scale to show the extent of the particle distribution. {\bf Bottom:} Phase space $v_r-r$ projection. The vertical dashed line corresponds to $R_{\rm 200m}$.}
    \label{fig:orbinf_split_example}
\end{figure*}

Indeed, it has long been known that the phase-space structure of matter around halos is bimodal \cite[e.g.][]{diemand_08}.  By using particle orbit information, it is possible to distinguish orbiting from infalling particles in the vicinity of a halo \citep{diemer2022a, diemer2022b}.  We illustrate this decomposition in Figure~\ref{fig:orbinf_split_example}, as achieved using the algorithm proposed by \citet{GarciaSalazar2022}, henceforth referred to as Paper I.  There, we classified every dark matter particle within $5\ \hMpc$ of halos identified with \rockstar\  \citep{Behroozi2013} as either orbiting or infalling based on: 1) the time at which the particle was first accreted; and 2) the particle's radial velocity, averaged over one dynamical time.  Orbiting particles are accreted early on, and have an average radial velocity close to zero, whereas infalling particles have only been recently accreted, and their average radial velocity is large and negative. We will refer to the orbiting/infall split based on these criteria as the \textit{accretion-time} criterion. 

Figure~\ref{fig:orbinf_split_example} shows the mass distribution and phase-space density of the most massive halo in the Cold Dark Matter (CDM) simulation by \citet{banerjeeetal20}, split into its orbiting and infall components.   The top row shows the particle density for all (left), orbiting (middle), and infalling (right) particles, split according to the accretion-time criterion.  The bottom row shows the corresponding phase-space plots for each of the particle populations.  We can see that the accretion-time criterion successfully decomposes the bimodal phase space distribution into its two dynamically distinct components.  Using the particle tracking code {\sc Sparta}, one may achieve similar decomposition by splitting particles according to whether or not they have experienced their first pericentric passage \citep[][]{diemer2022a, diemer2022b}.

Motivated by these observations, in Paper I we proposed to define a dark matter halo as the collection of all particles orbiting their self-generated potential.   This definition has several significant benefits:
\begin{enumerate}
\setlength\itemsep{0pt}
    \item It is parameter free.
    \item It respects particle dynamics (Figure~\ref{fig:orbinf_split_example}).
    \item It does not impose halo exclusion.
    \item It does not impose spherical symmetry on intrinsically triaxial halos \citep{Jing_2000, knebe_wiessner_2006}.
    \item Dynamical halos have a finite spatial extent \citep{diemer2022a,diemer2022b}  characterized by a well-defined halo radius \citep{diemer25, Shields2025} that is nearly redshift independent at fixed peak height \citep{diemer2022a} when expressed in units of $R_{200m}$.
    \item The mass function, bias, and clustering of matter around dynamical halos are simpler to model than those derived using standard spherical overdensity definitions \cite{GarciaSalazar2022,salazar2023}.
\end{enumerate}
In short, \it dynamical halos simplify large-scale structure modeling \rm \citep[see also][]{diemer25}. 

The physical motivation for the orbiting/infall decomposition advocated in these recent works was further strengthened in recent work by \citet{strawberry}, who used the boosted potential introduced by \citet{boosted} to split particles in the vicinity of dark matter halos according to whether they are \textit{bound} or \textit{unbound}, finding that infalling particles become bound between their first pericentric passage and their first apocentric passage.  That is, the bound/unbound criterion is very similar to the various orbiting/infall criteria described above.

\begin{table*}[!t]
    \centering
    \begin{tabular}{|c|c|c|c|c|c|c|c|}
        \hline
        Name & $\Omega_{\rm m}$ & $\Omega_{\rm b}$ & $h$ & $n_s$ & $\sigma_8$ & $L_{\rm box}~(h^{-1}{\rm Mpc})$ & $m_p~(h^{-1}{\rm M}_{\odot})$ \\ \hline
        Banerjee & $0.3$ & $0.049$ & $0.70$ & $0.96$ & $0.85$ & $1000$ & $7.75 \times 10^{10}$  \\ \hline
        \multirow{6}{*}{Erebos WMAP} & \multirow{6}{*}{$0.27$} & \multirow{6}{*}{$0.0469$} & \multirow{6}{*}{$0.70$} & \multirow{6}{*}{$0.95$} & \multirow{6}{*}{$0.82$} & $2000$ & $5.58 \times 10^{11}$  \\ 
         & & & & & & $1000$ & $6.98\times10^{10}$ \\
         & & & & & & $500$ & $8.72\times10^{9}$ \\
         & & & & & & $250$ & $1.09\times10^{9}$ \\
         & & & & & & $125$ & $1.36\times10^{8}$ \\
         & & & & & & $62.5$ & $1.70\times10^{7}$ \\ \hline
        \multirow{3}{*}{Erebos Planck} & \multirow{3}{*}{$0.32$} & \multirow{3}{*}{$0.0491$} & \multirow{3}{*}{$0.67$} & \multirow{3}{*}{$0.9624$} & \multirow{3}{*}{$0.834$} & $500$ & $1.03\times 10^{10}$  \\ 
         & & & & & & $250$ & $1.29\times10^{9}$ \\
         & & & & & & $125$ & $1.62\times10^{8}$ \\ \hline
        Quijote Latin Hypercube & $[0.1, 0.5]$ & $[0.03, 0.07]$ & $[0.5, 0.9]$ & $[0.8, 1.2]$ & $[0.6, 1.0]$ & $1000$ & -- \\ \hline
        Quijote Fiducial Planck & $0.3175$ & $0.049$ & $0.6711$ & $0.9624$ & $0.834$ & $1000$ & $8.21 \times 10^{10}$ \\ \hline
    \end{tabular}
    \caption{Cosmological parameters of all simulations used.  All boxes contain $1024^3$ particles.}
    \label{tab:cosmologies}
\end{table*}

Unfortunately, the benefits of these more physically motivated approaches to halo structure come at a cost: distinguishing orbiting from infalling particles requires accessing the trajectory of every particle in the vicinity of each halo. This is a numerically costly operation that requires dedicated software tools \citep[e.g.][]{sparta1}, making finding dynamical halos 
numerically intensive. Machine learning can be used to speed up classifications, but that still demands at least two time steps \cite{Vladimir2025}.  To address this difficulty, we set out to identify a simple, single-snapshot criterion capable of distinguishing between orbiting and infalling particles, with the goal of identifying a fast dynamical halo finding algorithm.  

Our starting point is the intuition that orbiting particles should be roughly virialized, so their kinetic energy should be roughly equal to half their potential energy ($T\approx -\tfrac{1}{2}V$).  Indeed, \citet{strawberry} find bound particles very nearly satisfy the virial equilibrium condition. By contrast, infalling particles started a large distance away from the halo, and therefore have nearly zero energy, implying $T\approx -V$.  That is to say, \it orbiting particles are expected to have lower kinetic energy than infalling particles at the same radius. \rm Here, we demonstrate that this basic intuition holds, and we exploit this feature to produce a new code called \oasis\ for Orbiting-mass ASsIgnment Scheme.\footnote{\url{https://github.com/baylihayes/oasis}} \oasis\ post-processes \rockstar\ halo catalogs to generate dynamical halo catalogs from single-snapshot simulation data.

The layout of our paper is as follows.  We present the simulation data used in this work in Section~\ref{sec:data}.  Next, we describe the basis of the \oasis\ algorithm in Section \ref{sec:algorithm}. In Section~\ref{sec:validation} we validate the \oasis\ algorithm by comparing against the accretion-based dynamical halo catalog presented in Paper I and the {\sc Sparta} algorithm, finding excellent agreement with both. Finally, as an example of the kind of questions we can now answer with \oasis, in Section~\ref{sec:massfunction} we characterize the cosmology-dependent mass function of dynamical halos using the Quijote simulation suite \citep{Quijote_sims} and two sets of Erebos simulations \cite{diemer2013a, diemerkravtsov14, diemer2015}.  We summarize our results and present some discussion in Section \ref{sec:discussion}.

%% file: sections/data.tex
\section{Simulation Data and Halo catalog}
\label{sec:data}

\begin{figure*}
    \centering
    \includegraphics[width=0.8\linewidth]{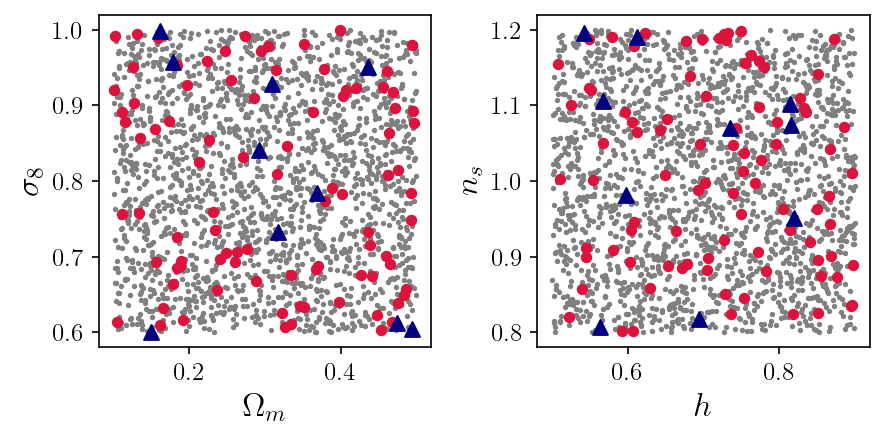}
    \caption{Latin hypercube cosmological parameter sample in the Quijote simulations (black points) and the sub-sample of simulations used in this work (red and blue points). The red points are Quijote simulations that were only analyzed at $z=0$, while the 10 blue triangular points are Quijote simulations that were analyzed at $z=0, 0.5, 1,2,$ and $3$.}
    \label{fig:lh_sample}
\end{figure*}

This work relies on a variety of simulations with flat $\Lambda$CDM cosmologies. All simulations contain $1024^3$ particles.  For the most part, we restrict our analysis to $z=0$ snapshots, though we also calibrate and test \oasis\ at a variety of different redshifts.  The simulations we use differ in the precise values of the cosmological parameters, as well as the box size.  For all simulations, the original halo catalog used in the simulation was created with \rockstar\ \citep{Behroozi2013}. In Section~\ref{sec:algorithm}, we use \it all \rm structures found by \rockstar\ as \textit{seeds} for our algorithm, regardless of whether they are labeled as parent halos or subhalos. Therefore, for the remainder of the discussion, we will sometimes refer to this input \rockstar\ catalog as the seed catalog.  We briefly summarize each of the simulations we use in our work both below and in Table \ref{tab:cosmologies}.

{\bf Banerjee Simulation:} The Banerjee et al.~\citep{banerjeeetal20} simulation is a $1\ \hGpc$ box with $\Omega_{\rm m}=0.3$ and $\sigma_8=0.85$ run using \gadget~\citep{Springel2005}.  This was the simulation we used in Paper I to motivate the definition of dynamical halos.  

{\bf Erebos Simulations:} We use two sets of Erebos simulations \cite{diemer2013a, diemerkravtsov14, diemer2015} described in \citet{diemer2015}. The first set uses the same cosmology as that of the Bolshoi simulation \cite{klypin_dark_2011}, and is consistent with WMAP7 \cite{wmap7}. We use six boxes of different sizes for this cosmology, ranging from a $62.5\ \hMpc$ box to a $2000\ \hMpc$ box, enabling us to probe a broad range of halo masses.  The second set of simulations adopted a Planck-like cosmology \cite{planck2013} and consists of three boxes of sizes $125\ \hMpc$, $250\ \hMpc$, and $500\ \hMpc$.

{\bf Quijote Simulations:} The Quijote suite of simulations (82,000 boxes) spans a wide range of varying cosmologies, box sizes, and particle resolutions \citep{Quijote_sims}. We use the Latin hypercube sub-sample of boxes to explore the cosmological dependence of our results.  This sub-sample of boxes varies $\Omega_{\rm m}$, $\Omega_{\rm b}$, $h$, $n_s$ and $\sigma_8$, while keeping the neutrino mass fixed to $M_\nu=0.0\ {\rm eV}$, and the dark energy equation of state to $w=-1$. We further reduce the Latin hypercube sample down to 100 simulations, which we select by maximizing the distance between any two cosmologies.  Specifically, we define the distance $d_{ij}^2$ between any two simulations via
\begin{equation}
    d_{ij}^2 = \sum_{p\in\theta}\left(\dfrac{p_{i}- p_{j}}{w_{p}}\right)^2,
\end{equation}
where $\theta=\{\Omega_{\rm m},\Omega_{\rm b}, h, n_s, \sigma_8\}$, and $w_p$ is the width of the parameter interval (see Table \ref{tab:cosmologies}). We assign a distance $d_i^2$ to each cosmology as the minimum distance of cosmology $i$ to all other cosmologies $j$, $d_i^2\equiv \min(d_{ij}^2)$.  We select our sub-sample of simulation boxes by iteratively selecting the cosmology with the largest distance among the remainder of the parent sample. Figure~\ref{fig:lh_sample} shows the distribution of the 2,000 Quijote cosmologies in parameter space (black points) and the sub-sample of 100 cosmologies we selected (red and blue points).  Both red and blue points were analyzed at $z=0$, while only the blue points were analyzed at $z=0,0.5,1,2,$ and $3$. For comparison to the $z=0$ Latin hypercube boxes in Section~\ref{sec:mf_mass_dependence}, we also use 100 Planck (Fiducial) cosmology boxes of the same size and particle number as the Latin hypercube sample.


%% file: sections/algorithm.tex
\section{Algorithm}
\label{sec:algorithm}

This section describes the \oasis\ post-processing algorithm used to generate dynamical halo catalogs from simulations.  Sections~\ref{sec:ke_space} through \ref{sec:ke_cut_cosmo} describe how we differentiate between orbiting and infalling particles on the basis of their kinetic energies.  This initial classification is refined around halo substructures as described in Section~\ref{sec:substructure}. Finally, Section~\ref{sec:percolation} describes how we percolate the input \rockstar\ halo catalog using these two algorithms to arrive at our final dynamical halo catalog.

\subsection{Energy-Based Particle Classification}
\label{sec:ke_space}

\begin{figure*}
    \centering
    \includegraphics[width=\linewidth]{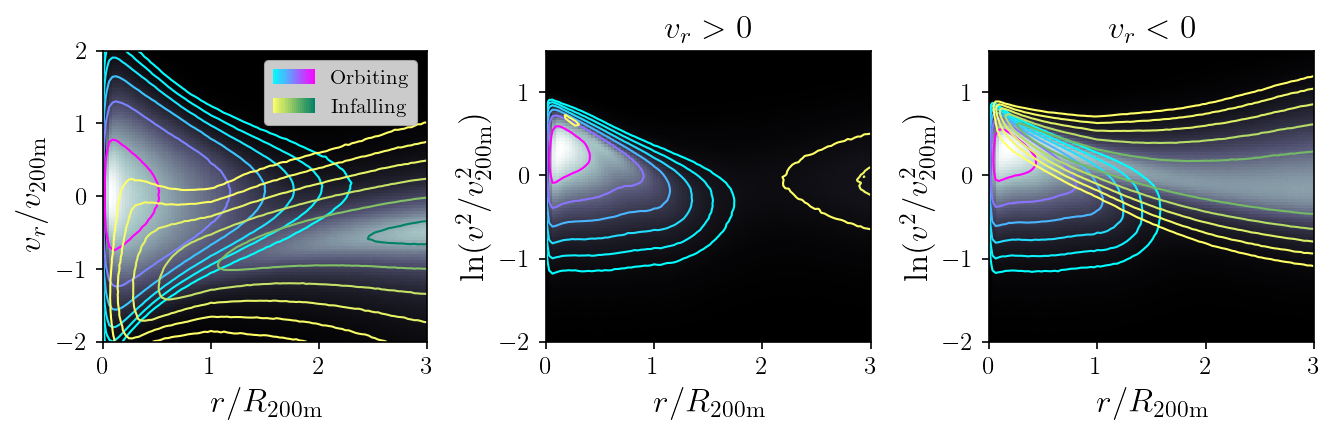}
    \caption{The distribution of orbiting and infalling particles in phase-space (left), and in the space of kinetic energy versus radius.  The latter is subdivided according to whether the particles have positive (middle) or negative (right) radial velocities.  The colored contours show the particle density after splitting the particles into their orbiting (blue to magenta) and infall (yellow to green) components using the accretion-time algorithm of Paper I. As expected, infalling particles have higher kinetic energies than orbiting particles.}  
    \label{fig:at_cuts}
\end{figure*}

We wish to establish an energy-based criterion that enables us to distinguish orbiting from infalling particles.  The left panel in Figure~\ref{fig:at_cuts} shows the phase-space distribution of orbiting and infalling particles around halos in the Banerjee et al. simulation.  For this figure, the particles were classified as either orbiting or infalling using the accretion-time criterion from Paper I. To account for a broad range of halo masses, radii are quoted in units of $R_{\rm 200m}$, and velocities in units of $v_{\rm 200m}$, where 
\begin{align}
    M_{\rm 200m}&=\frac{4\pi}{3}\Delta\ \rho_{\rm m}R_{\rm 200m}^3, \label{eq:r200m}\\
    v_{\rm 200m}&=\sqrt{\frac{GM_{\rm 200m}}{R_{\rm 200m}}},
\end{align}
where $\Delta=200$.\footnote{Note the algorithm in our code release expects the radii of the particles and $R_{\rm 200m}$ to be in comoving units, making our input $v_{\rm 200m}$ comoving as well; \oasis\ converts it to physical units internally.} Below, we discuss further how we treat $R_{\rm 200m}$ and $v_{\rm 200m}$ when making our classification. As can be seen from Figure~\ref{fig:at_cuts}, the phase-space distribution of orbiting and infalling particles overlap in radius: there is no radial cut that can cleanly distinguish orbiting from infalling particles.

Our intuition is that ``orbiting'' particles can be roughly interpreted as bound and virialized, while ``infalling'' particles are roughly unbound \citep[see also][]{strawberry}.  Consequently, we anticipate orbiting and infalling particles to separate in the space of kinetic energy vs. radius. Because the kinetic energy of particles rapidly increases as they reach the pericenter, we use the log of the kinetic energy, which varies roughly linearly with radius. The central and right panels of Figure~\ref{fig:at_cuts} show the distribution of orbiting and infalling particles in this space, as classified by accretion time, split by whether the particles have positive (central panel) or negative (right panel) radial velocities.  Because particle mass is constant, we ignore it when making this plot. While some particle overlap remains, our basic intuition holds true: orbiting particles have lower kinetic energies than infalling particles at the same radius.  Moreover, infalling particles almost always have instantaneous negative radial velocities.  Motivated by this figure, we propose to redefine orbiting particles using linear cuts in the (log) kinetic energy vs. radius space
\begin{equation}\label{eq:vr_cut}
    \ln\left(\frac{v^2}{v_{\rm 200m}^2}\right) = m(x-x_0)+b,
\end{equation}
where $x=r/R_{\rm 200m}$ and $v$ is the particle velocity relative to the halo velocity.\footnote{Following the GADGET-2 format, \oasis\ expects the input particle velocities to be stored as $v_{\rm part, sim} = v_{\rm part, physical} / \sqrt{a}$. Likewise, following \rockstar\ format, \oasis\ expects the velocity of the halo seed to be in physical units.} 

We parameterize the amplitude of our linear cuts in terms of the value of the cut at a pivot point $x_0=0.5$. We will refer to this selection as the \it kinetic energy \rm criterion, which stands in contrast to the accretion-time criterion we used in Paper I.   In what follows, we describe how we can self-calibrate the amplitude and slope of the proposed cuts from the simulation data.  However, we note that in Section~\ref{sec:ke_cut_cosmo} we calibrate the resulting cuts as a function of cosmology. Consequently, the \oasis\ code allows the user to skip the self-calibration of the kinetic energy cuts in favor of the cosmology-calibrated cuts presented in Section~\ref{sec:ke_cut_cosmo}.

\begin{figure*}
    \includegraphics[width=0.8\linewidth]{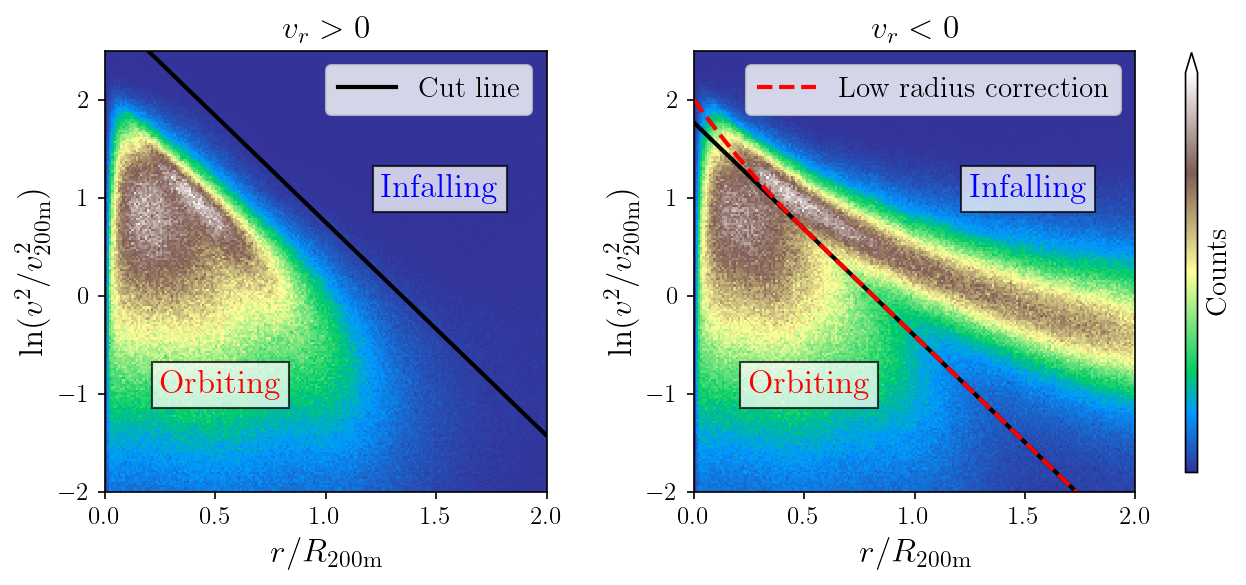}
    \caption{Particle density in the space of kinetic energy vs. radius for the 500 most massive isolated halos in the Banerjee et al. simulation, after restricting ourselves to halos that dominate their environment (see text).  We consider particles with positive (left) and negative (right) radial velocities separately. The black solid line shows the self-calibrated kinetic energy cut used to separate orbiting from infalling particles in the simulation: particles above the line are labeled as infalling, while particles below the line are labeled as orbiting.  Note the kinetic energy cuts differ between the two panels. The red dashed line shows our modification to the kinetic energy cuts necessary to ``catch'' the upper-left tip of the orbiting particle distribution.  We refer to this modification as the low-radius correction, motivated by the work in \citep{Vladimir2025}.}
    \label{fig:calibration}
\end{figure*}

To self-calibrate the kinetic energy cut that defines the kinetic energy criterion, we started by identifying the sub-sample of \rockstar\ halos that dominate their environment, meaning the next most massive halo within a $2R_{\rm 200m}$ radius has a mass no larger than 20\% of that of the central halo.  We then selected the most massive halos from this sample, and determined the particle density in the kinetic energy vs. radius space, for both positive and negative radial velocity particles. The number of halos selected for the fitting varies with particle mass and simulation box size. We used 500 halos for large (1,000$\ \hMpc$ or higher) boxes.  For the smaller Erebos boxes, we selected 150 halos for the $250$ and $500\ \hMpc$ boxes, and $50$ halos for the smallest boxes.  These choices reflect the fact that the number of particles per halo decreases very quickly for the large boxes, but slowly for the small boxes, so the latter require fewer halos to enable accurate calibration of the kinetic energy cuts.

With this selection in hand, we determine the slope of the linear kinetic energy cut. To do so, we examine the distribution $P_y(y)$ of log kinetic energy, $y = \ln(v^2/v_{\rm 200m}^2)$,  for positive radial velocity particles at a set of different radii within the range $r/R_{\rm 200m} \in[0.2,0.5]$. For each radius, we identify a characteristic kinetic energy at which the log kinetic energy distribution has the steepest decline, i.e. the value at which the Gaussian-smoothed estimate of $-dP_y(y)/dy$ reaches its maximum. We fit the resulting characteristic values of $y$ as a function of $r/R_{\rm 200m}$ with a line and adopt the slope of this line as the slope of the proposed kinetic energy cut for both positive and negative radial velocities. The amplitude at the pivot point for negative radial velocities is then selected by the value of $b$ which maximizes the mean squared perpendicular distance to all particles within a narrow band of half-width $w=0.05$ centered on the cut line.  This is roughly equivalent to minimizing the number of particles in the band, but we found this criterion improved numerical stability. Finally, for positive radial velocities the pivot point amplitude is chosen such that 99.5\% of all positive radial velocity particles fall below the proposed cut. We have done so in order to exclude rare fly-by particles. Note that modest changes to this amplitude have a negligible impact on the resulting halo masses.

In addition, particles sufficiently far from the central halo cannot be orbiting, so we label all particles with radii $r\geq 2R_{\rm 200m}$ as infalling, irrespective of their radial velocities. In principle, there can be a tiny amount of orbiting mass beyond this radial cut, but its fractional contribution to the orbiting mass is below 1\% \citep{sparta1}.

The black solid lines in Figure~\ref{fig:calibration} show the kinetic energy cuts derived using the above algorithm for the Banerjee et al. simulation.  Particles above the line are labeled as infalling, while particles below the line are labeled as orbiting. As desired, the proposed kinetic energy cuts effectively separate orbiting from infalling particles. 

We emphasize that, in deriving our proposed kinetic energy cuts, we rely on the radius $R_{\rm 200m}$ using all particles.  However, when \it applying \rm these cuts, we make use of the ``bound'' spherical overdensity mass and radius $M_{\rm 200b}$ and $R_{\rm 200b}$ reported by \rockstar.\footnote{Throughout this work, the subscript $\rm 200b$ denotes a spherical overdensity of 200 times the mean matter density computed using only gravitationally bound particles, as in \rockstar, while $\rm 200m$ denotes the same overdensity computed using all particles. Quantities with the $\rm 200b$ subscript, such as the velocity scale $v_{\rm 200b}$, are computed from these bound masses and radii.} The reason is that using $R_{\rm 200b}$ improves halo percolation: for small mass halos in the immediate vicinity of massive halos, $R_{\rm 200b}$ is a much better estimate of the size of the halo than $R_{\rm 200m}$. Consequently, while the cuts are calibrated with $R_{\rm 200m}$ and $v_{\rm 200m}$, we apply Equation~\eqref{eq:vr_cut} with $x = r/R_{\rm 200b}$ and $v_{\rm 200m}$ replaced by $v_{\rm 200b}$. The same substitution applies to Equations~\eqref{eq:vr_cut_neg} and~\eqref{eq:vr_cut_cosmo}.

\subsection{Low-Radius Correction}
\label{sec:sparta}

\citet{Vladimir2025} performed a detailed comparison of the kinetic energy cuts proposed here with the orbiting/infalling classification obtained in Erebos using {\sc Sparta} \citep{sparta1}.  They found the linear cuts adopted here fail to adequately capture the high kinetic energy tails of orbiting particles with negative radial velocities at small radii.  However, this can be fixed by curving the kinetic energy cut upwards as we move towards small radii, as illustrated by the dashed-red line in the right panel of Figure~\ref{fig:calibration}.  This improved kinetic energy cut differs from our original cut only to the left of the pivot point $0.5R_{\rm 200b}$.  Specifically, our modified cut takes the form
\begin{equation}\label{eq:vr_cut_neg}
    \ln\left( \frac{v^2}{v^2_{\rm 200b}}\right) = \begin{cases}
    \alpha x^2 +\beta x+\gamma & x\leqslant x_0, \\
    {\rm Equation~\eqref{eq:vr_cut}} & x>x_0.
    \end{cases}
\end{equation}
The parameters $\alpha$ and $\beta$ governing the cut at small radii are chosen to ensure smoothness of the cut at the pivot scale $r=0.5\ R_{\rm 200b}$. This results in
\begin{align}
\alpha = \frac{\gamma - b_{\rm neg} + m_{\rm neg}\,x_0}{x_0^2}, \\
\beta  = m_{\rm neg} - 2\alpha x_0,
\end{align}
where $m_{\rm neg}$ and $b_{\rm neg}$ are the slope and pivot amplitude corresponding to $v_r<0$. That is, $\alpha$ and $\beta$ are not independently fit. $\gamma$ is an additional free parameter that we set to $\gamma=2$ using the data from \cite{Vladimir2025}. The resulting cut is the red dashed line in the right panel of Figure~\ref{fig:calibration}.  We apply this correction irrespective of how the original linear kinetic energy cuts are derived, i.e. via self-calibration or using our cosmology-dependent cuts.  Our final cut in kinetic energy space is therefore Equation~\eqref{eq:vr_cut} for positive radial velocities and Equation~\eqref{eq:vr_cut_neg} for negative radial velocities.

\subsection{Cosmology, Redshift, and Box-Size Dependence of the Kinetic Energy Criterion}
\label{sec:ke_cut_cosmo}

\begin{figure*}
    \includegraphics[width=\linewidth]{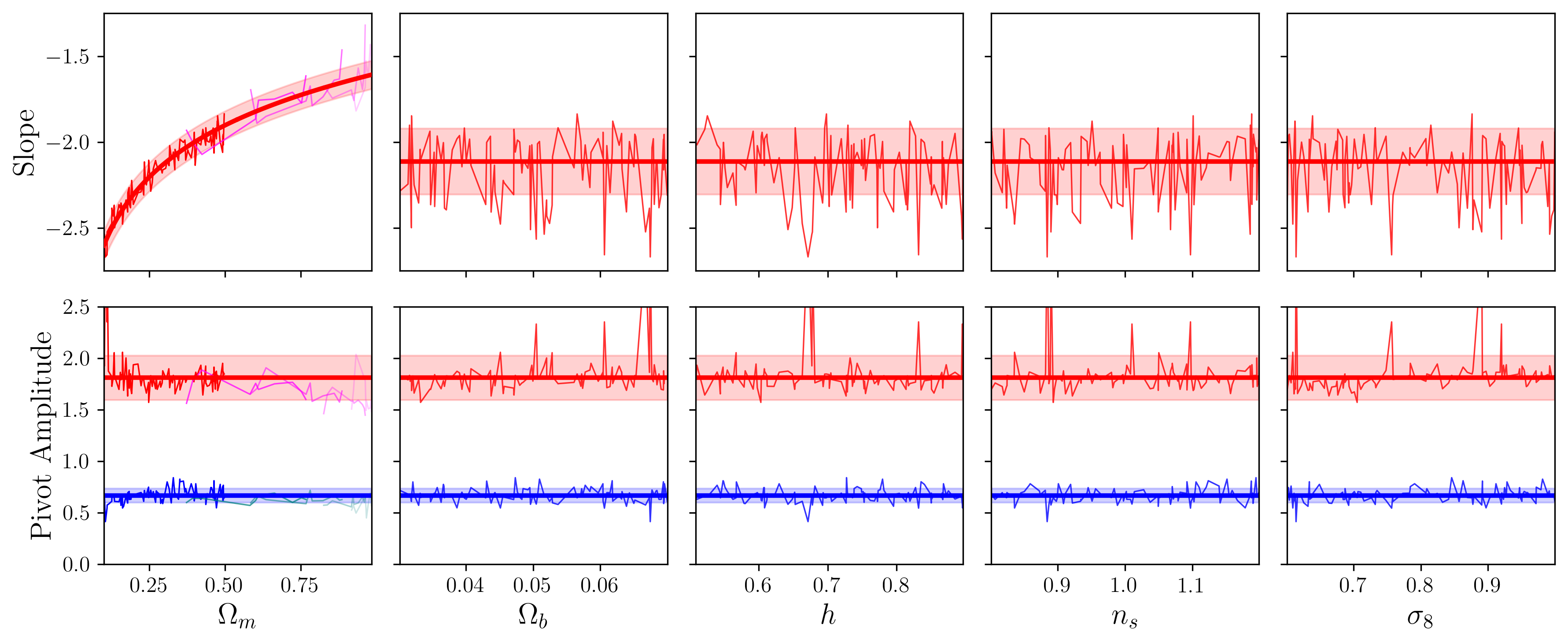}
    \caption{Slope (top row) and pivot amplitude (bottom row) values from the calibration procedure of the kinetic energy cuts derived from the Latin hypercube sample boxes from the Quijote simulation suite. Each column corresponds to a different cosmological parameter. The red and blue colors correspond to particles with positive and negative radial velocities, respectively.  The data points are shown as joined lines to ease visualization of the trend. The solid lines are fits to the data and the shaded regions correspond to the rms dispersion around the fit curve. The purple and cyan curves in the leftmost panels correspond to 10 Quijote simulations that were also analyzed at $z=0.5,1,2,$ and $3$. The decreasing opacity corresponds to a higher redshift.} 
    \label{fig:calib_quijote}
\end{figure*}

\begin{figure*}
    \includegraphics[width=0.9\linewidth]{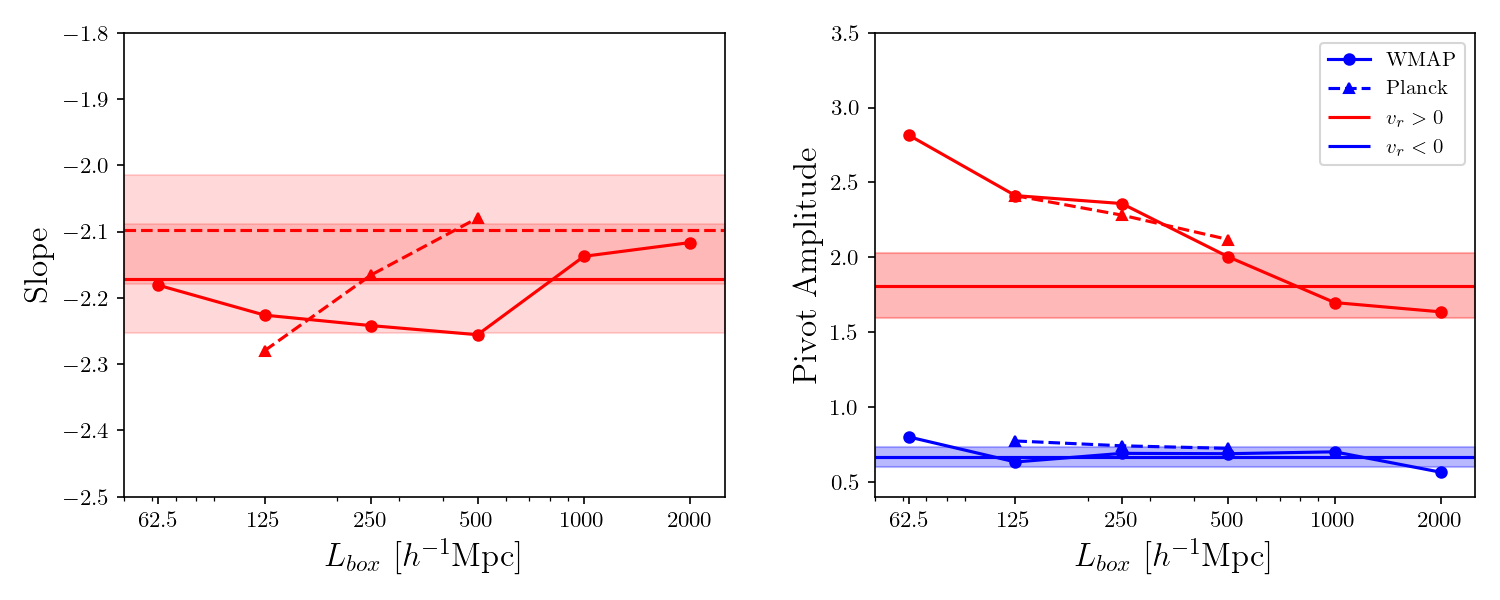}
    \caption{Dependence of the slope (left) and pivot amplitude (right) on box size in the Erebos simulation suite, as labeled. Red and blue colors correspond to positive and negative radial velocity particles, respectively. The solid horizontal lines correspond to those shown in Figure~\ref{fig:calib_quijote}, i.e., they are the Quijote-calibrated cuts.  The bands around the solid lines are the rms dispersion of the parameters in the Quijote simulations. The large differences between the red band and the red curves reflect the fact that fly-by particles are more common in the type of low-mass halos sampled using small boxes.  However, these differences have little impact on the mass of dynamical halos since fly-by particles are rare.
    }
    \label{fig:calib_erebos}
\end{figure*}

To test how the output of the kinetic energy self-calibration algorithm depends on box size and cosmology, we applied the above algorithm to the Erebos and Quijote simulations described in Section~\ref{sec:data}. Remarkably, we find that at $z=0$ the pivot amplitude and slope of our cuts are independent of both cosmology (Figure~\ref{fig:calib_quijote}) and mass (or box-size, Figure~\ref{fig:calib_erebos}), with one exception: \textit{the slope of the kinetic energy cuts depends logarithmically on the matter density of the Universe}. 

To extend our algorithm to higher redshifts, we ran \oasis\ on 10 simulations (blue triangles in Figure~\ref{fig:lh_sample}) at $z=0,0.5,1,2,$~and~3 to self-calibrate the kinetic energy cut as a function of redshift.  Figure~\ref{fig:calib_quijote} demonstrates that the trend in slope of the cut as a function of $\Omegam$ extends smoothly to high redshift, provided the $x$-axis is $\Omegam(z)$. The resulting best fit model for our kinetic energy cut is
\begin{equation}\label{eq:vr_cut_cosmo}
    \ln\left( \left.\frac{v^2}{v^2_{\rm 200b}}\right|\ \Omega_m\right) = m(\Omega_m)(x-x_0)+b(\Omega_m),
\end{equation}

where $m$ is the slope, $b$ is the amplitude at the pivot $x_0=0.5$, and $x=r/R_{\rm 200b}$. The slope of the cut is
\begin{equation}
    m(\Omegam) = -1.903 + 0.435\ln\left(\frac{\Omega_{\rm m}(z)}{0.5}\right).
\end{equation}
The amplitude terms depend on whether a particle has a positive or a negative radial velocity as follows:
\begin{align}
    b_{\rm pos} & = 1.809, \\
    b_{\rm neg} & = 0.666.
\end{align}
As we can see in the leftmost panels of Figure~\ref{fig:calib_quijote}, the parameterization above accurately describes the resulting kinetic energy cuts. That is, when expressed in terms of $\Omegam(z)$, the kinetic energy cuts appear to be universal.

To test whether the cuts we calibrated using Quijote apply to lower mass halos, we further compare our Quijote calibration to the self-calibrated cuts from the Erebos simulations in Figure~\ref{fig:calib_erebos}.  The bands around the lines in that figure characterize the rms fluctuations of the parameters in the Quijote simulations, highlighting the consistency in the cuts between Erebos and Quijote for the $v_r<0$ particles. However, Figure~\ref{fig:calib_erebos} also reveals a mismatch in the pivot amplitude and slope of the self-calibrated kinetic energy cuts between Erebos and Quijote for $v_r>0$ particles.  Fortunately, this cut is only made to remove fly-by particles (see left panel in Figure~\ref{fig:calibration}), which tend to be more frequent for the low-mass halos found in small boxes.  Since fly-by particles are rare, the differences seen in Figure~\ref{fig:calib_erebos} have little physical consequence for the final orbiting/infall classification of particles in the vicinity of dark matter halos.  Note that users of \oasis\ can choose either our cosmology-dependent calibrated cuts, or self-calibration from the simulation data as described above.

\subsection{Substructure Corrections}
\label{sec:substructure}

\begin{figure*}
    \centering
    \includegraphics[width=0.95\linewidth]{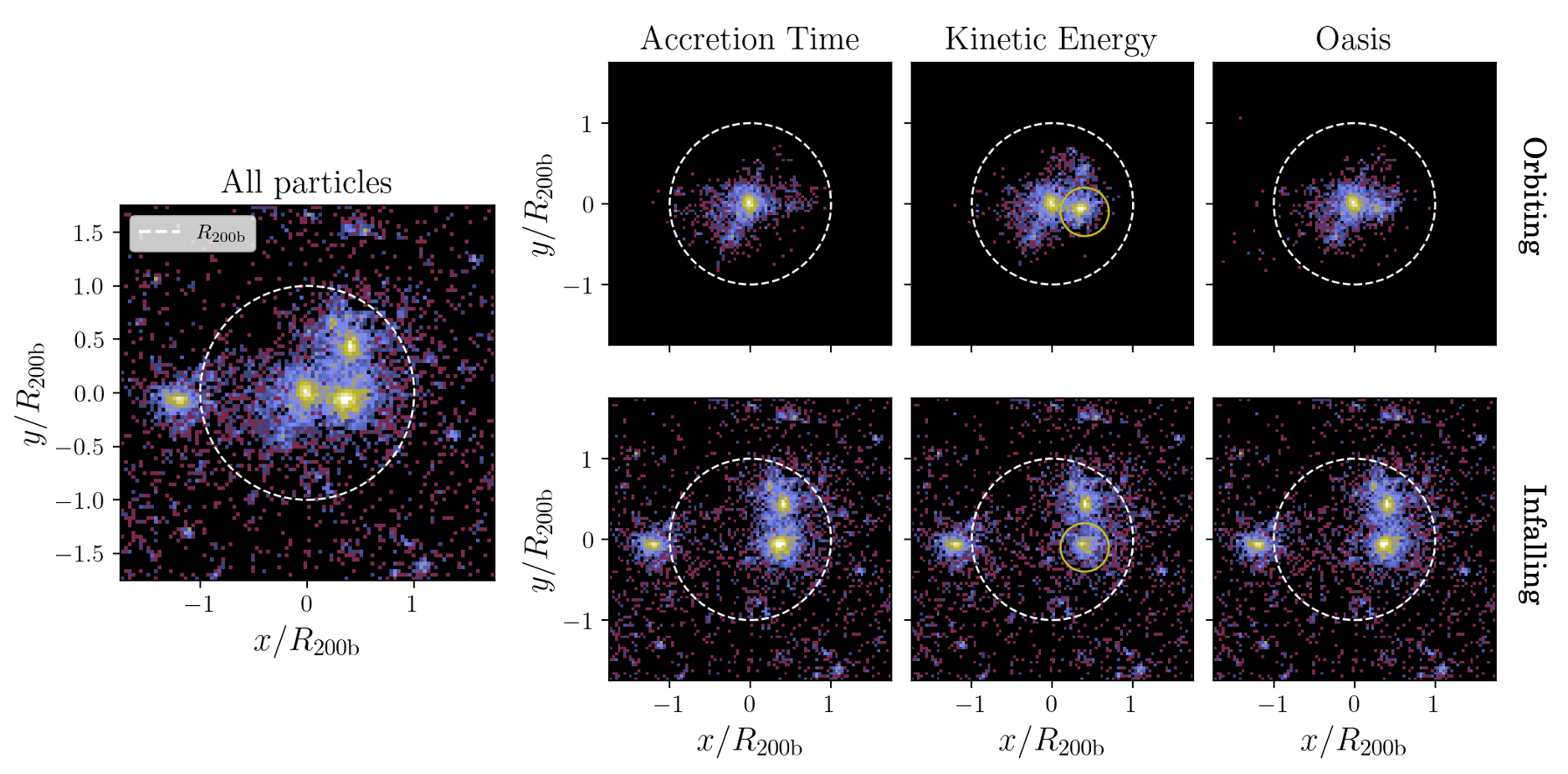}
    \caption{\textbf{Left:} Sample halo of mass $M_{\rm 200m}=6.2\times 10^{14}\hMsun$. \textbf{Right:} The same halo, but split into orbiting (top row) and infalling (bottom row) particles according to the accretion-time criterion (left), kinetic energy cut (middle), and \oasis\ (right). Only a fraction of particles within the region encircled by the yellow contour are labeled as orbiting based on their kinetic energy. After applying our substructure correction step, \oasis\ labels all particles associated with this structure as infalling.}
    \label{fig:substructure_correction}
\end{figure*}

Unfortunately, the proposed kinetic energy cuts above fail to adequately account for the impact of substructure in the particle density.  This is illustrated in Figure~\ref{fig:substructure_correction}, where we show the orbiting (top row) and infalling (bottom row) particle density field around a sample halo with $M_{\rm 200m}=6.2\times 10^{14}\hMsun$ in the Banerjee et al. simulation.  The left column shows the orbiting/infall particle decomposition according to the accretion-time criterion from Paper I, while the middle column shows the particle decomposition obtained from applying the kinetic energy cuts described in the previous section.  We can see that when using the accretion-time criterion, nearly all particles of every substructure in the vicinity of the central halo are classified in the same way, so that each substructure appears only in one of the two maps (orbiting or infalling).  By contrast, when applying a kinetic energy cut (middle column), substructures can show up in \it both \rm maps, as highlighted by the substructure in the yellow circle. This is clearly undesirable.

To address this, after applying our kinetic energy cuts, we re-classify every particle within some radius $\Rball$ of 
a candidate halo to ensure that all such particles have the same label, be it orbiting or infalling.\footnote{\oasis\ uses all structures identified by \rockstar\ as candidate halos, and percolates through them as detailed below (Section~\ref{sec:percolation}).}  The label applied to all particles is determined by the fraction $\forb$ of the particles within this ``ball'' that are classified as orbiting.  All that remains is to specify: 1) how the radius $\Rball$ is determined; and 2) how to set the threshold $\forb$ used to determine whether particles within the ball are orbiting or infalling.  As we will see below, setting $\forb=1/2$ is suboptimal.

Consider a small \rockstar\ halo --- it can be a parent halo or a subhalo according to \rockstar\ --- of mass $m_{\rm 200b}$ at a radius $R$ from a more massive central halo of mass $M_{\rm 200b}$. We wish to determine  if this object is a parent halo that has not yet started to orbit the central halo, or whether it is a subhalo of the more massive central halo.  If the radial separation between the two halos is larger than $2R_{\rm 200b}$ of the central halo, the smaller halo is automatically labeled as infalling, and all its particles are infalling.  For halos with radial separations lower than $2R_{\rm 200b}$, we proceed as follows. Consider the line connecting the central halo to the smaller halo, and recall that this radial separation is $R$.  We identify the point along this line at which the density of the smaller halo is expected to match the density of the central halo.  The distance between this point and the smaller halo is the radius $\Rball$.  That is, $\Rball$ satisfies the equation
\begin{equation}
    \rho_{\rm cen}(R-R_{\rm ball}) = \rho_{\rm sub}(R_{\rm ball}).
\end{equation}
To find $\Rball$, we model the matter density profile of both structures with a Navarro, Frenk, and White \citep{NFW} profile, where the concentration of the profiles is obtained using the scale radius reported by \rockstar\ for each halo.  For small halos that are far from the central halo, it is possible for $\Rball$ to be larger than the radius $r_{\rm 200b}$ of the smaller halo, which is undesirable. Consequently, we further set
\begin{equation}
    \Rball={\rm min}\left\{ \Rball, r_{\rm 200b} \right\}.
\end{equation}
With this definition, $\Rball$ corresponds to the radius at which particles within the ball are more likely to belong to the substructure than to the central halo.  However, even within this ball there may be fast moving particles that are not bound to the substructure.  That is, particles assigned to the substructure must not only fall within the radius $\Rball$, they must also have a low velocity relative to the substructure.  Therefore, we extend the ``ball'' to velocity space, and demand the relative velocity of particles in the ball to satisfy
\begin{equation}
    v \leq v_{\rm ball} \equiv 2v_{\rm 200b},    
\end{equation}
where $v_{\rm 200b}$ is the velocity scale associated with the mass $m_{\rm 200b}$ of the smaller halo. 

Now that we have determined which particles ``belong'' to the substructure, we decide how to reclassify them.  To do so, we calculate the fraction $\forb$ of particles in the ball that have been classified as ``orbiting'' the larger central halo.  Naively, one might expect that if $\forb>1/2$, then we should label all particles in the ball as orbiting.  However, this is not correct: if the substructure is far from the halo, it is more likely for the substructure to be infalling.  This information should act as a prior that raises the $\forb$ threshold required to ``convert'' all particles within the ball into orbiting particles.  Indeed, as the distance between structures $R$ increases, the $\forb$ threshold should approach unity.  Consequently, we set our $\forb$ threshold via
\begin{equation}\label{eq:forb_threshold}
    f_{\rm orb}\geqslant\max\left\{\frac{1}{2}\ ,\ 1-\exp\left[-\left(\frac{R}{r_{\rm 200b}}\right)^{2}\right]\right\}.
\end{equation}
From visual inspection, adopting this radius-dependent threshold performs better than a constant $\forb=1/2$ threshold.

The rightmost column of Figure~\ref{fig:substructure_correction} illustrates the impact of applying the substructure correction algorithm in our classification of orbiting and infalling particles for our sample halo. Recall the central column corresponds to the particle classification resulting from the kinetic energy criterion.  Comparing the central and rightmost columns, we see that the highlighted structure (yellow circle) that was ``split'' by the kinetic energy criterion now appears as an infalling parent halo after the application of the substructure correction algorithm, in agreement with the accretion-time based classification.  It is, of course, possible for us to misclassify a substructure using this algorithm, labeling it as infalling when it is in fact orbiting or vice versa.  However, we found it preferable to identify sub-structures as either entirely orbiting or infalling rather than splitting the substructure into an orbiting half and an infalling half. For accurate structure classification and/or detailed halo dynamics, one is better served with dedicated particle tracking algorithms such as {\sc Sparta}. 

\subsection{Halo Percolation}
\label{sec:percolation}

We construct a dynamical halo catalog by taking the collection of all structures found by \rockstar\ on a given snapshot, and assigning an orbiting mass to each structure, as well as determining whether the structure is a parent halo or a subhalo.  We refer to this process as percolation.  In detail, we first rank-order the \rockstar\ halos by their $M_{\rm 200b}$ mass.  Starting from the most massive halo, we classify all particles within $2R_{\rm 200b}$ of the halo as either orbiting or infalling using the kinetic energy criterion, and then apply the substructure correction of Section~\ref{sec:substructure} to every less massive seed within its volume that has not already been tagged as a substructure of a more massive halo. Repeating this for every seed yields a preliminary orbiting mass, $M_{\rm orb}$, for each halo.\footnote{During this initial classification, \oasis\ does not remove particles labeled as orbiting from consideration for subsequent halos, so a single particle may contribute to the preliminary $M_{\rm orb}$ of several halos. These repeated assignments are resolved by the percolation step below. This is more numerically efficient than removing particles sequentially during the initial classification.} We then rank-order the halos by $M_{\rm orb}$ and percolate them in descending order: particles and substructures already assigned to a halo with a larger $M_{\rm orb}$ are removed from all subsequent halos, ensuring that every particle and substructure orbits at most one halo, and halos left with fewer than a specified minimum number of orbiting particles are discarded. The end result is a dynamical halo catalog in which every halo is defined as the collection of its orbiting particles.


Though we have described our algorithm as sequential, in practice we parallelize it to improve performance. To do so, we divide the simulation box into cubic cells that are processed independently and in parallel. Each cell includes a surrounding padding region so that particles and seeds associated with halos near the cell edge are included. We construct a dynamical halo catalog for each cell, retaining only halos whose centers reside within the cell's core region to avoid duplicate halo entries, and combine the cell catalogs into the dynamical halo catalog of the full simulation box. For simulations too large to hold in memory, the code can additionally group cells into tiles, each built from chunks of the input particle and seed data with an overlapping buffer region, and process the tiles separately. All results in this paper were produced without the tiling option. 


One critical feature of the resulting halo catalog is that \it dynamical halos are not subject to halo exclusion. \rm This can be seen in the lower-right panel of Figure~\ref{fig:substructure_correction}: the ``obvious by-eye'' substructures in this panel are all identified as parent halos by our algorithm, despite some of these structures falling within the $R_{\rm 200b}$ radius of the central halo. This is an important advantage over traditional halo-finding algorithms, since halo exclusion effects can lead to difficult-to-model features in the halo power spectrum \citep{Baldauf2013}.


%% file: sections/validation.tex
\section{Validation}
\label{sec:validation}

To validate the \oasis\ algorithm, we generated the \oasis\ dynamical halo catalogs for the Banerjee et al. and the 1000 $h^{-1}$ Mpc WMAP Erebos simulations, and compared them to the dynamical halo catalogs from Paper I and {\sc Sparta}, respectively.

For the Banerjee et al. simulation, we go down to parent halos with at least 50 orbiting particles. Note that the Paper I catalog is incomplete below a mass threshold of $4\times 10^{13}\ \hmsun$, or $\sim 500$ particles (see Paper I for details). For the Erebos simulation, we compare against the {\sc Sparta} catalogs, which go down to halos with 200 particles.  To allow for downscatter, we restrict our analysis to \oasis\ halos of mass $M_{\rm orb}\geq 10^{13.25}\ \hmsun$ , corresponding to $\approx 250$ particles.

\begin{figure*}
    \centering
     \includegraphics[width=0.9\linewidth]{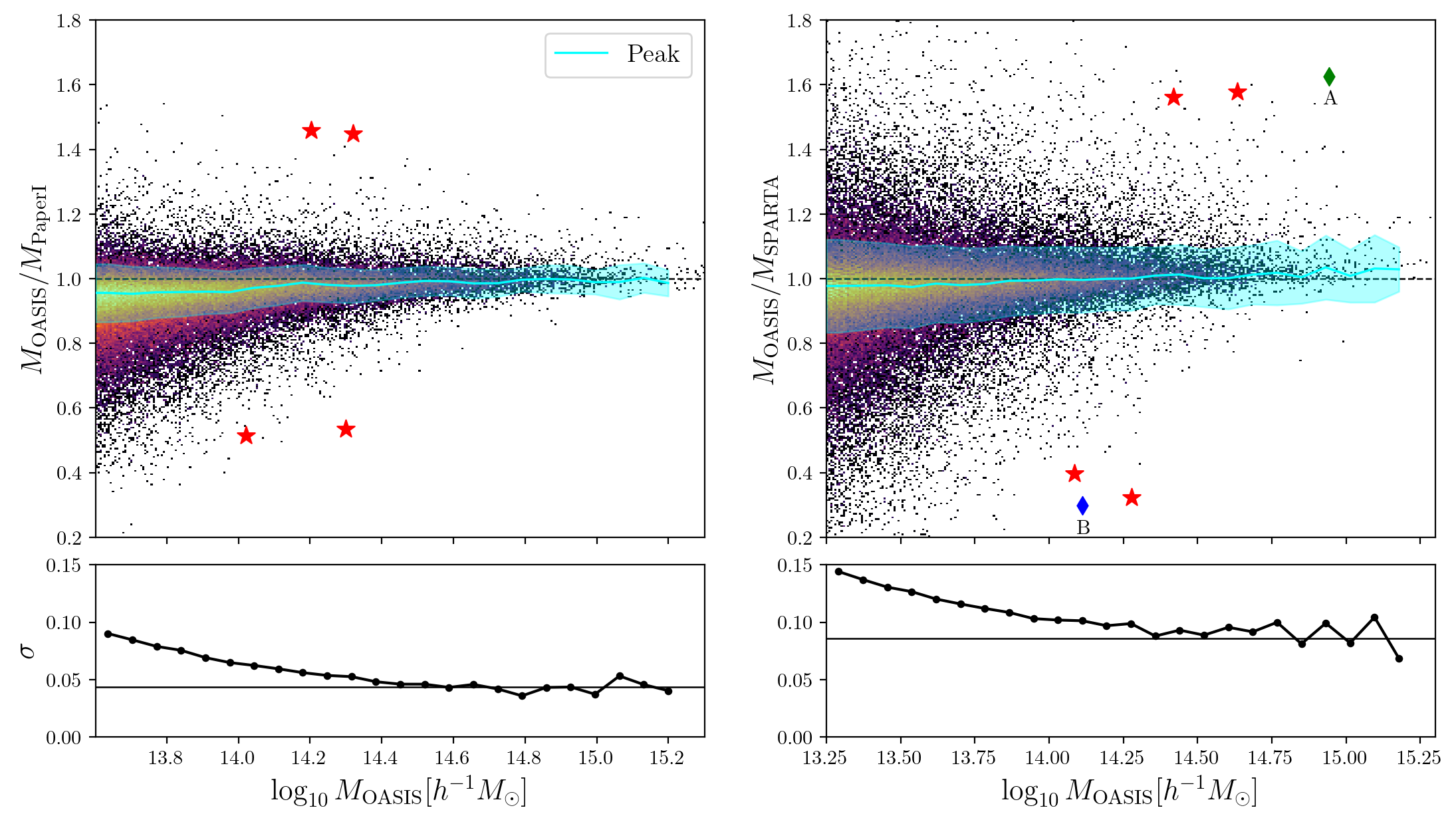}
    \caption{\textbf{Top:} Ratio of the \oasis\ to accretion time (left) and {\sc Sparta}~(right) orbiting mass of individual halos as a function of \oasis\ orbiting mass. The color scale is logarithmic to accentuate the dispersion at the tails of the distribution. The cyan solid line is the location of the peak of the distribution as a function of mass, and the shaded region is the standard deviation. The red-starred points correspond to the four outlier halos for Paper I and {\sc Sparta} shown in  Figures~\ref{fig:TA_pos_comparison} and~\ref{fig:SPARTA_pos_comparison}, respectively. The green (Halo A) and blue (Halo B) diamonds are the two halos shown in Figure~\ref{fig:sparta_special_case}. \textbf{Bottom:} Standard deviation as a function of mass. The solid black line is the asymptotic value at high masses of $0.044$ and $0.086$, respectively.}
    \label{fig:comparison2}
\end{figure*}

\subsection{Comparison of Individual Halo Masses}
\label{sec:mass_comparison}

We compare the orbiting masses of halos calculated by \oasis\ to those we obtained in Paper I, and those obtained using the {\sc Sparta} classifier.  Starting with Paper I, the top left panel of Figure~\ref{fig:comparison2} shows the ratio of orbiting masses obtained using \oasis\ and the accretion time criterion. The mass ratio is shown as a function of the \oasis\ orbiting mass.  Note that since all halos are generated from \rockstar\ halos as seeds, there is no ambiguity in halo matching (halos share halo ids). 

The peak of the distribution of mass ratios in Figure~\ref{fig:comparison2} (cyan solid line) occurs at $\approx 0.957$, though this peak ratio increases steadily towards unity with increasing mass. Unsurprisingly, the standard deviation in the mass ratio (cyan band) increases with decreasing number of particles, but asymptotes to a value of $0.044$ at the high-mass end.  We interpret this value as the intrinsic scatter between the orbiting masses of the two algorithms.  The logarithmic color scale denotes the density of halos at each point, and highlights the emergence of low-ratio tails in the low-mass end due to differences in halo percolation.  The near-unity peak mass ratio, and the small scatter about this value, provide strong evidence that \oasis\ is working as intended.

The right panels of  Figure~\ref{fig:comparison2} perform a similar comparison using the Erebos simulation. The peak ratio is $\approx0.979$ with a typical scatter of $\approx 8.6\%$.  These results further strengthen our conclusion that \oasis\ is working as intended.

\begin{figure*}
    \centering
    \includegraphics[width=0.8\linewidth]{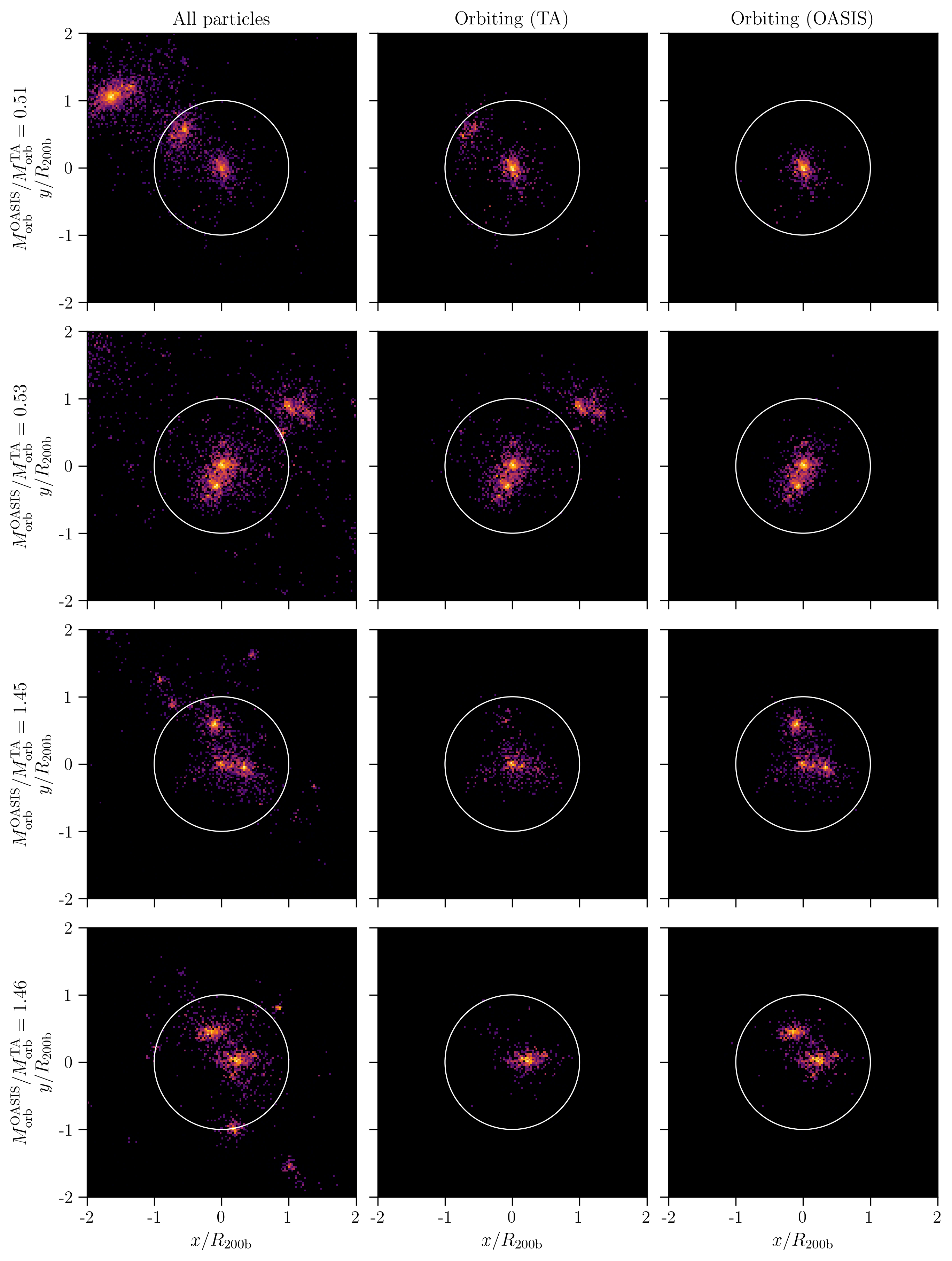}
    \caption{Four sample Banerjee et al. halos shown as the red-starred points in the top-left panel in Figure~\ref{fig:comparison2}. Each row corresponds to a different halo. The \oasis /Paper I orbiting mass ratio for each halo is indicated on the vertical axis label. The halos shown have unusually low (top two rows) or high (bottom two rows) mass ratios.  The white circle is the radius $R_{\rm 200b}$ for each halo. \textbf{Left:} All (infalling and orbiting) particles associated with each halo. \textbf{Middle:} Orbiting particles given by the selection scheme of Paper I (TA, or accretion-time). \textbf{Right:} Orbiting particles given by \oasis.}
    \label{fig:TA_pos_comparison}
\end{figure*}

\begin{figure*}
    \centering
    \includegraphics[width=0.8\linewidth]{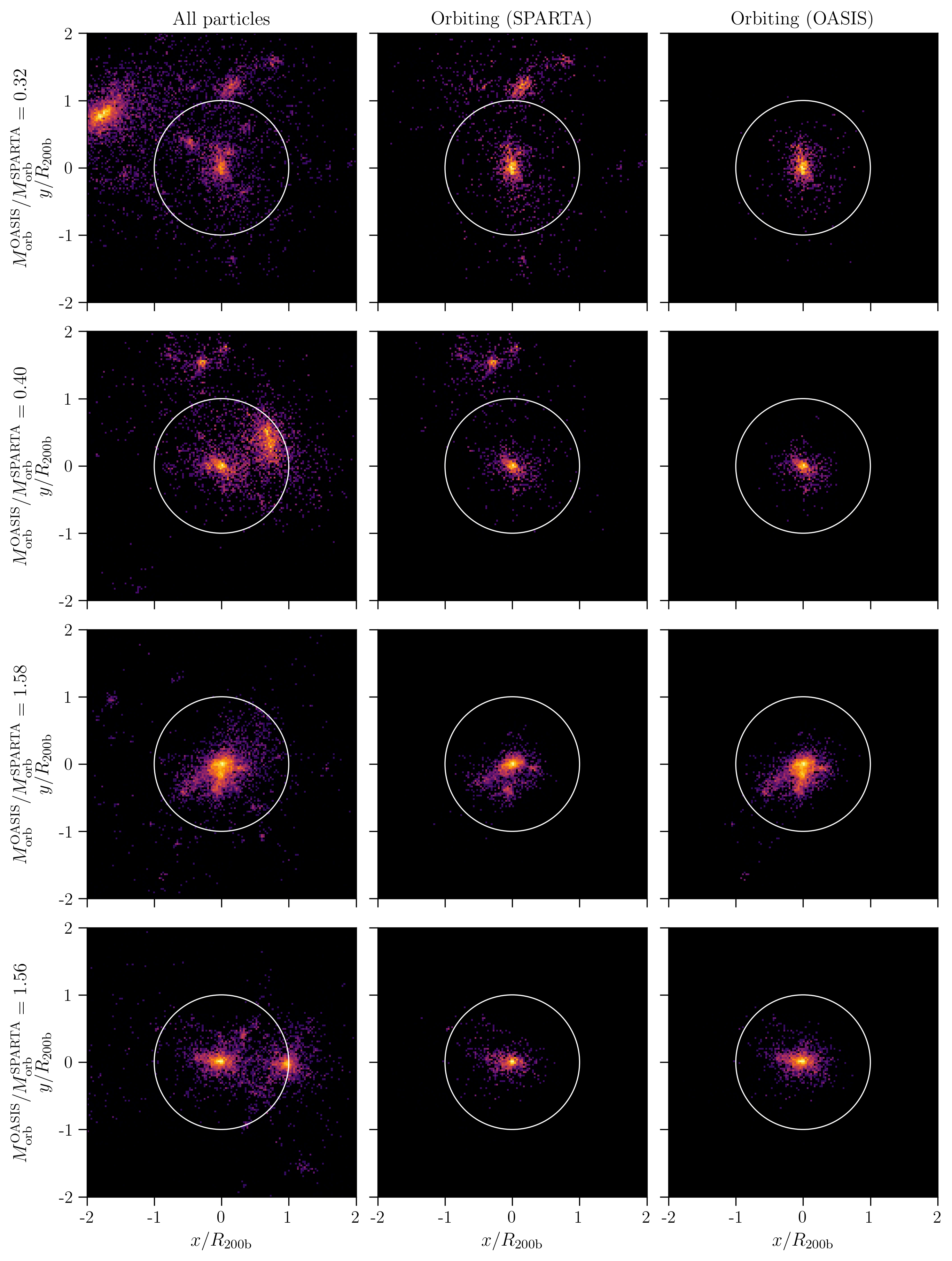}
    \caption{Same format as Figure~\ref{fig:TA_pos_comparison}, but with the four Erebos halos found as the red-starred points in the top-right plot of Figure~\ref{fig:comparison2}. The orbiting masses obtained using \oasis\ are compared with those obtained by {\sc Sparta}.}
    \label{fig:SPARTA_pos_comparison}
\end{figure*}

\begin{figure*}
    \centering
    \includegraphics[width=0.8\linewidth]{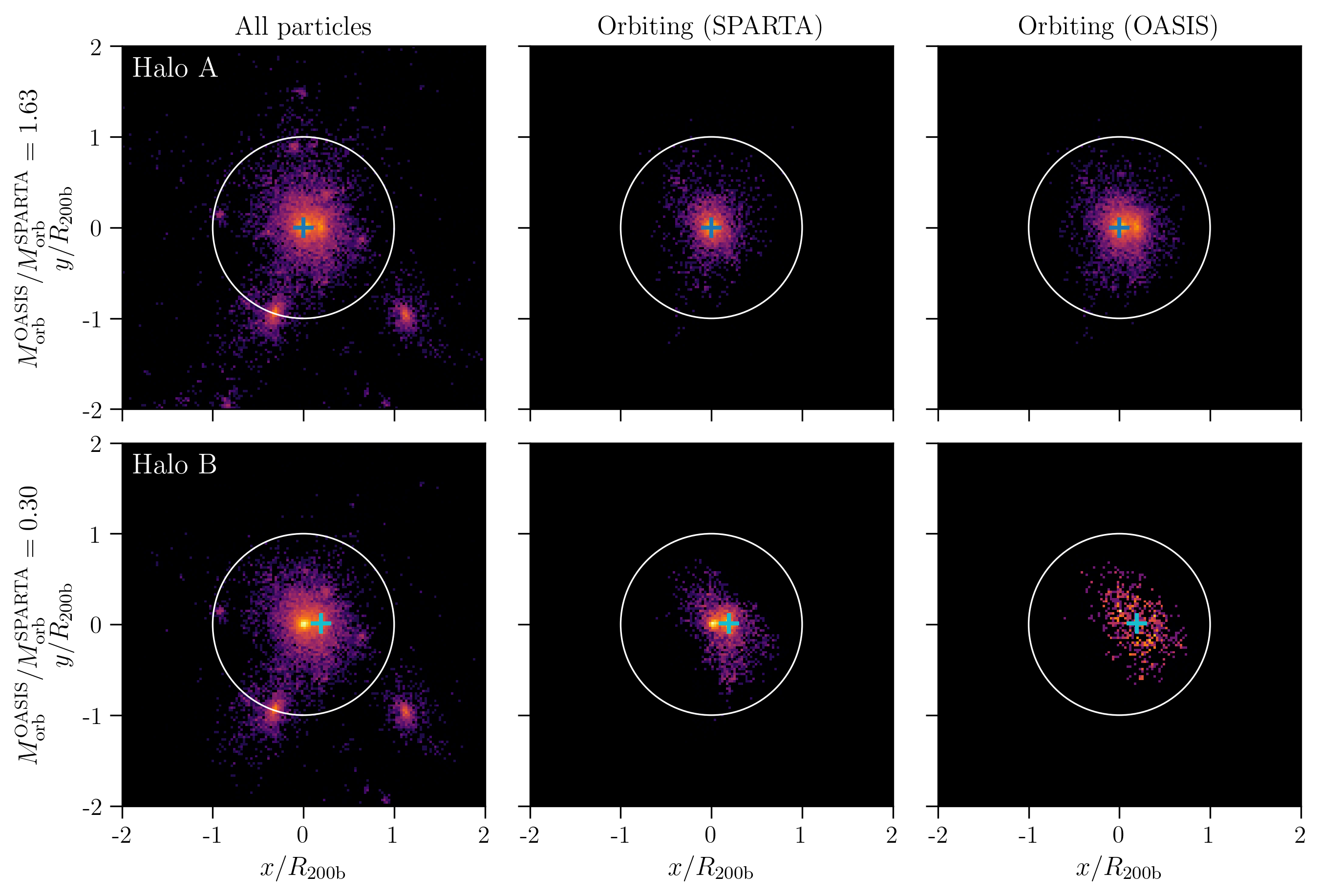}
    \caption{A halo system that demonstrates how percolation can lead to unusually low \oasis/{\sc Sparta} mass ratios.  The two halos above are halos A and B highlighted as the green and blue diamonds in Figure~\ref{fig:comparison2}. We see that \oasis\ identifies the structure on the right (Halo B) as a subhalo of the halo on the left (Halo A). \oasis\ therefore assigns the particles of Halo B to Halo A, so when we get to Halo B, it has nearly no particles left, leading to a very low \oasis/{\sc Sparta} mass ratio.
    }
    \label{fig:sparta_special_case}
\end{figure*}

To better understand the relative differences between \oasis\ and the orbiting/infalling selection schemes of Paper I and {\sc Sparta}, we visually inspected halos with $\Morb\geq 10^{14}\ \hMsun$ exhibiting unusually high or unusually low mass ratios, as well as several halos with mass ratios close to unity.  The conclusions from the eye-balling can be summarized as follows:
\begin{itemize}
    \item Halos with mass ratios close to unity tend to be isolated, with most particles having unambiguous classifications.
    \item Major mergers can lead to unusually high or unusually low mass ratios when the two algorithms differ in the classification of the merging structure.  For such major mergers, it is difficult to determine which algorithm is correct by eye, but we anticipate {\sc Sparta} will most often be correct, as it uses much more information in its analysis.  
    \item Related to the previous failure mode, large differences can also arise due to differences in percolation: a parent halo in one algorithm can be percolated away in the other algorithm.
    \item Extreme low mass ratios can also arise from halos that are badly miscentered in the input \rockstar\ catalog.  We leave the development of improved halo centering to future work.
\end{itemize}

As an illustration, we have selected four halos --- two with a high \oasis/reference mass ratio, and two with a low mass ratio --- for both the Paper I and {\sc Sparta} analyses.  The selected halos are shown as red stars in Figure~\ref{fig:comparison2}.  Projections of the particles for each of these halos are shown in Figures~\ref{fig:TA_pos_comparison} (Paper I) and~\ref{fig:SPARTA_pos_comparison} ({\sc Sparta}).  For each figure, the top two rows correspond to low \oasis/reference mass ratios, while the bottom two rows correspond to high-mass ratios. We can see that in all cases the central halo is undergoing significant merging activity, making it difficult to unambiguously classify neighboring structures as orbiting or infalling.

The green and blue diamonds in Figure~\ref{fig:comparison2} correspond to two halos that help illustrate how percolation can exacerbate these difficulties, labeled Halo A and Halo B. Figure~\ref{fig:sparta_special_case} shows that \oasis\ finds Halo B to be a substructure of Halo A. Most of the particles of Halo B are therefore tagged as orbiting particles of Halo A, and are not available for Halo B: when \oasis\ looks for a halo at the position of Halo B, it only finds a diffuse cloud of unbound particles with very little mass.  By contrast, {\sc Sparta} tags Halo B as an infalling halo, and its particles are not associated with Halo A.  Thus, the percolation step in \oasis\ in which the particles of Halo B get retagged as orbiting particles of Halo A is responsible for the unusually low \oasis/{\sc Sparta} mass ratio for Halo B. 

Overall, we consider the agreement between \oasis\ and both Paper I and {\sc Sparta} a success.  Specifically, the slight mass differences between the algorithms are a small price to pay for being able to quickly and efficiently generate dynamical halo catalogs from simulations.

\subsection{Mass Function Comparison}
\label{sec:MF_comparison}

Having compared the orbiting masses of individual halos, we now turn to comparing the mass function from Paper I to the mass function recovered with \oasis. Following convention, we plot the mass function in terms of the multiplicity function $f(\nu)$, where 
\begin{equation}\label{eq:mulitplicityfunction}
    f(\nu) = \dfrac{1}{\bar{\rho}_m}\dfrac{dn}{d\ln M}\left(\dfrac{d\nu}{dM}\right)^{-1}
\end{equation}
and
\begin{equation}\label{eq:peakheight}
    \nu = \dfrac{\delta_{\rm sc}(z)}{\sigma(M, z)}
\end{equation}
is the peak height associated with each halo.  For the purposes of this comparison, we have ignored the finite-box corrections necessary for a fully correct calculation of $\sigma(M,z)$ discussed in the next section.

\begin{figure}
    \centering
    \includegraphics[width=0.9\linewidth]{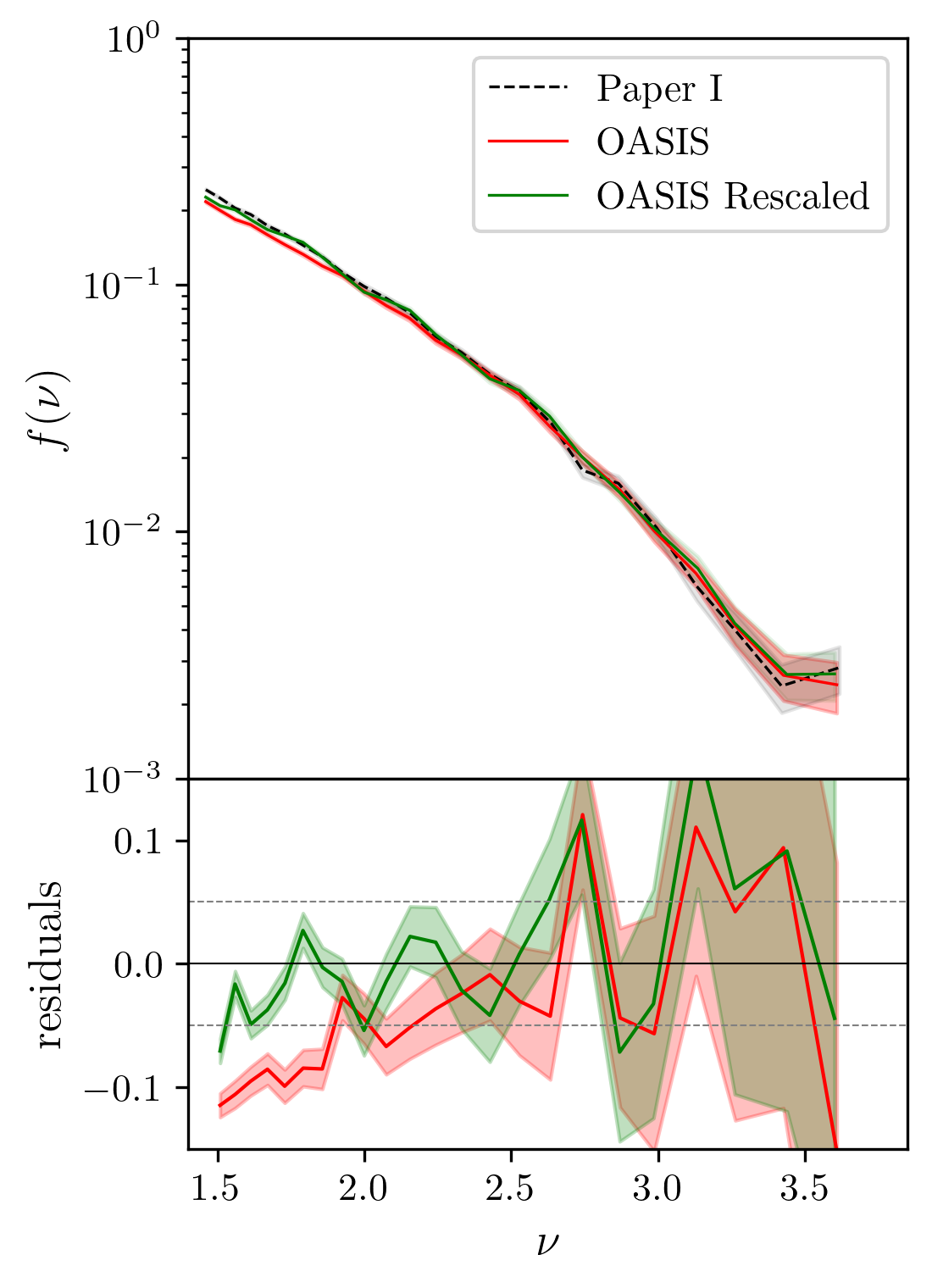}
    \caption{{\bf Top:} Multiplicity function of dynamical halos as a function of peak height for accretion time (black dashed line) and \oasis\ (solid lines). Red shows the original OASIS multiplicity function; green shows the result after dividing the orbiting masses by the mass-dependent peak ratios shown by the cyan curve in the upper-left panel of Figure~\ref{fig:comparison2}. The bands for the \oasis\ curves are jackknife errors. {\bf Bottom:} Log-ratio of the \oasis\ multiplicity functions to that of Paper I. The banded regions are the $\pm1\sigma$ uncertainties propagated from the \oasis\ multiplicity functions. The dashed horizontal lines indicate approximate $\pm 5\%$ deviation marks. Rescaling reduces the systematic negative offset at low peak heights (\(\nu\lesssim2.5\)).}
    \label{fig:comparison}
\end{figure}

Figure~\ref{fig:comparison}  shows that the \oasis\ mass function is somewhat tilted relative to that derived using the accretion time algorithm, though the relative differences are small, typically below 10\%. If the orbiting masses from \oasis\ are rescaled by $1/r$, with $r$ being the most likely mass ratio (cyan line) in the top left plot of Figure~\ref{fig:comparison2}, we see that the two mass functions are in excellent (better than 5\%) agreement.  That is, the differences in percolation between Paper I and \oasis\ have little impact on which parent halos we find, but the masses of those halos differ slightly between the two algorithms.


%% file: sections/massfunction.tex
\section{The Mass Function of Dynamical Halos}
\label{sec:massfunction}

Having demonstrated the efficacy of the \oasis\ algorithm, we wish to highlight the kind of analysis enabled by \oasis\ by considering the functional form and cosmological dependence of the mass function of dynamical halos. Throughout, we focus specifically on the multiplicity function $f(\nu)$ (see Section~\ref{sec:MF_comparison}).  Our goal in this section is to be illustrative rather than complete, and we anticipate we will present a more detailed analysis of the dynamical halo mass function in a future publication.

\subsection{The Functional Form of the Mass Function}
\label{sec:mass_function_model}

\begin{figure*}
    \centering
    \includegraphics[width=0.8\linewidth]{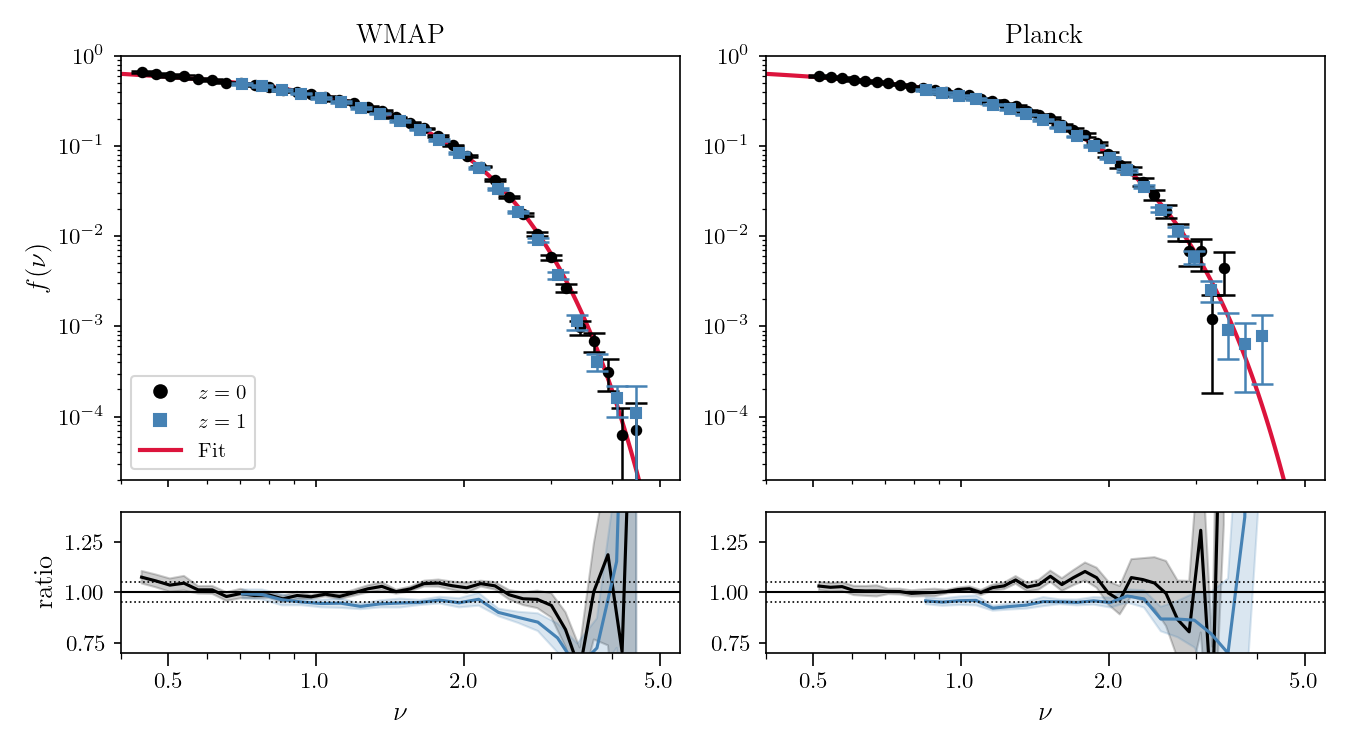}
    \caption{{\bf Top:} Inverse-variance-weighted multiplicity function for the Erebos WMAP7 (left) and Planck (right) boxes for $z = 0$ (black circles) and $1$ (blue squares). The best fit model for Equation~\eqref{eq:fnu} is shown as a red line and is a fit to all $z = 0$ boxes simultaneously. Error bars represent 1$\sigma$ uncertainties derived from the diagonal of the jackknife covariance matrices and combined using inverse variance weighting. {\bf Bottom:} Ratios of the data to the best fit model. Shaded regions show the corresponding $\pm 1\sigma$ uncertainties. Dotted horizontal lines are at the $\pm5\%$ deviation marks.}
    \label{fig:Erebos_mf}
\end{figure*}

Figure~\ref{fig:Erebos_mf} shows the mass function of dynamical halos derived from the Erebos simulations for redshifts $z =0$ and $1$ (points with error bars). To construct this mass function, we generate an \oasis\ halo catalog for each of the Erebos boxes, restricting the analysis to halos with 500 particles or more. The covariance matrix for each box is estimated using jackknife resampling.  We fit the data from all $z = 0$ boxes simultaneously and adopt a multiplicity model for the mass function of the form \citep[e.g.][]{shethtormen,Diemer2020}

\begin{equation}
    f(\nu) = A\left[1+b\nu^{-a}\right]\exp\left(-\dfrac{\nu^2}{2}\right),
    \label{eq:fnu}
\end{equation}
where
\begin{equation}
    A = \left[ \sqrt{\dfrac{\pi}{2}} + \dfrac{b\,\Gamma\left(\frac{1-a}{2}\right)}{2^{\frac{1+a}{2}}} \right]^{-1},
\end{equation}
and $\Gamma$ is the gamma function.  Note the parameter $a$ is restricted to the range $0\leqslant a<1$.  In addition to $a$ and $b$, the threshold for collapse $\dsc$ is treated as a fit parameter.  When $b=0$ this parameterization reduces to the standard Press--Schechter form \cite{press-schechter1974}. 

We fit the binned differential mass function by maximizing a Gaussian likelihood with the full jackknife covariance. We bin halos in logarithmic mass bins, extending the mass range up to the first empty bin above which there are no more halos. To prevent the covariance from becoming artificially small in sparsely populated bins, we add a diagonal variance term equal to one count per bin across all mass bins included in the fit. This also ensures a nonzero variance for the most massive, empty bin, where the jackknife variance vanishes. We also allow for an intrinsic fractional scatter $\epsilon$ along the diagonal for every mass bin, and fit for $\epsilon$ as part of the fit. The total covariance used in our fits is
\begin{equation}
    C_{ij} = C^{\rm JK}_{ij}
    + \delta_{ij}\left[s_i^2 + \epsilon^2 m_i^2\right],
\end{equation}
where $s_i = [V\,\Delta\ln M_i]^{-1}$ is the contribution to the differential mass function corresponding to one halo count in bin $i$, $V$ is the simulation volume, and $m_i$ is the predicted differential mass function in bin $i$. The parameter $\epsilon$ characterizes fractional scatter about the model beyond the jackknife uncertainties. We include the most massive empty bin to include the information that ``no halos are found in this bin'', but note that removing this term from the likelihood has little impact on our fits. When fitting the mass function data, we have accounted for the suppression in the number of high-mass halos in each box due to the lack of power above the box scale \citep{reed_etal07} by subtracting the box variance from $\sigma^2(M)$ \citep{lukic_etal07}. 

In order to properly constrain all mass function parameters, $f(\nu)$ must be sampled at both small and high peak height values.  We find that the range of $\nu$ values probed by the Planck Erebos boxes is not quite sufficient for this purpose, so we have chosen to simultaneously fit the Erebos WMAP and Planck boxes assuming the two cosmologies have identical $a$, $b$ and $\delta_{\rm sc}$ parameters. The resulting best fit parameters are summarized in Table~\ref{tab:mass_function_fit}, and correspond roughly to $a\approx0.8$, $b\approx0.1$, and $\dsc\approx1.54$ for both WMAP and Planck cosmologies.  This is lower than the canonical 1.68 value, but within the expected range of $1.52$--$1.68$ estimated by \citep{Shapiroetal99}. The best fit intrinsic scatter of our model is $1.4\%$.  Figure~\ref{fig:Erebos_mf} compares the inverse variance weighted mass function constructed using all Erebos boxes to our best fit model (red solid line).  

\begin{table}
    \centering
    \begin{tabular}{|l|l|r|}
        \hline
        Parameter & Description & Value \\\hline
        $a$ & Slope & $0.779\ \pm 0.023$  \\
        $b$ & Amplitude & $0.123\ \pm 0.012$ \\ 
        $\delta_{\rm sc}$ & Collapse threshold & $1.544\ \pm 0.003$ 
        \\\hline
        $\log_{10}\varepsilon$ & Intrinsic scatter & $-1.846\ \pm 0.059$  \\\hline
    \end{tabular}
    \caption{Best fit parameters of the multiplicity function in Equation~\eqref{eq:fnu} for the Erebos boxes in both WMAP and Planck cosmologies. Quoted uncertainties are the mean of the upper and lower 68\% profile-likelihood intervals.}
    \label{tab:mass_function_fit}
\end{table}

In Paper I, we found that the mass function of dynamical halos could be adequately described using a Press--Schechter multiplicity function with a slowly varying threshold for collapse that spanned the range of reasonable collapse values $\dsc \in [1.52,1.68]$. However, this is no longer the case when we sample the mass function across a broad range of halo masses, as demonstrated by the Erebos data. Indeed, if we demand that the halo multiplicity function be described by an evolving collapse threshold $\dsc$, we can solve for $\dsc$ as a function of peak height from the Erebos data using the standard Press--Schechter formula
\begin{equation}\label{eq:ps}
    f(\nu) = \sqrt{\frac{2}{\pi}} \exp\left(-\dfrac{\nu^2}{2}\right).
\end{equation}
Doing so, we find that low mass halos from the Erebos simulation would require $\dsc \gtrsim 5$, a physically implausible value.  However, if we restrict ourselves to peak heights comparable to those of Paper I, we find similar results to those reported in that work.

\subsection{The Cosmology Dependence of the Halo Mass Function}
\label{sec:mf_mass_dependence}

As an illustrative example of the type of work enabled by \oasis, we now turn to the cosmology dependence of the mass function, which we characterize using the Quijote simulation suite. For this analysis, we do not fit the data to the model given in Equation~\eqref{eq:fnu}; instead, we compare it to 100 Quijote boxes with a Planck, or fiducial, cosmology at the same box size and particle number as the Latin hypercube boxes, enabling us to look for cosmology dependence in a non-parametric way.  Here, we consider the sharply limited question of whether the mass function at $z=0$ has a universal form as we vary $\Omega_{\rm m}$, $n_{\rm s}$, and $\sigma_8$.  A complete characterization of the cosmology dependence of the dynamical halo mass function will be presented in future work.

\begin{figure*}
    \centering
    \includegraphics[width=1.0\linewidth]{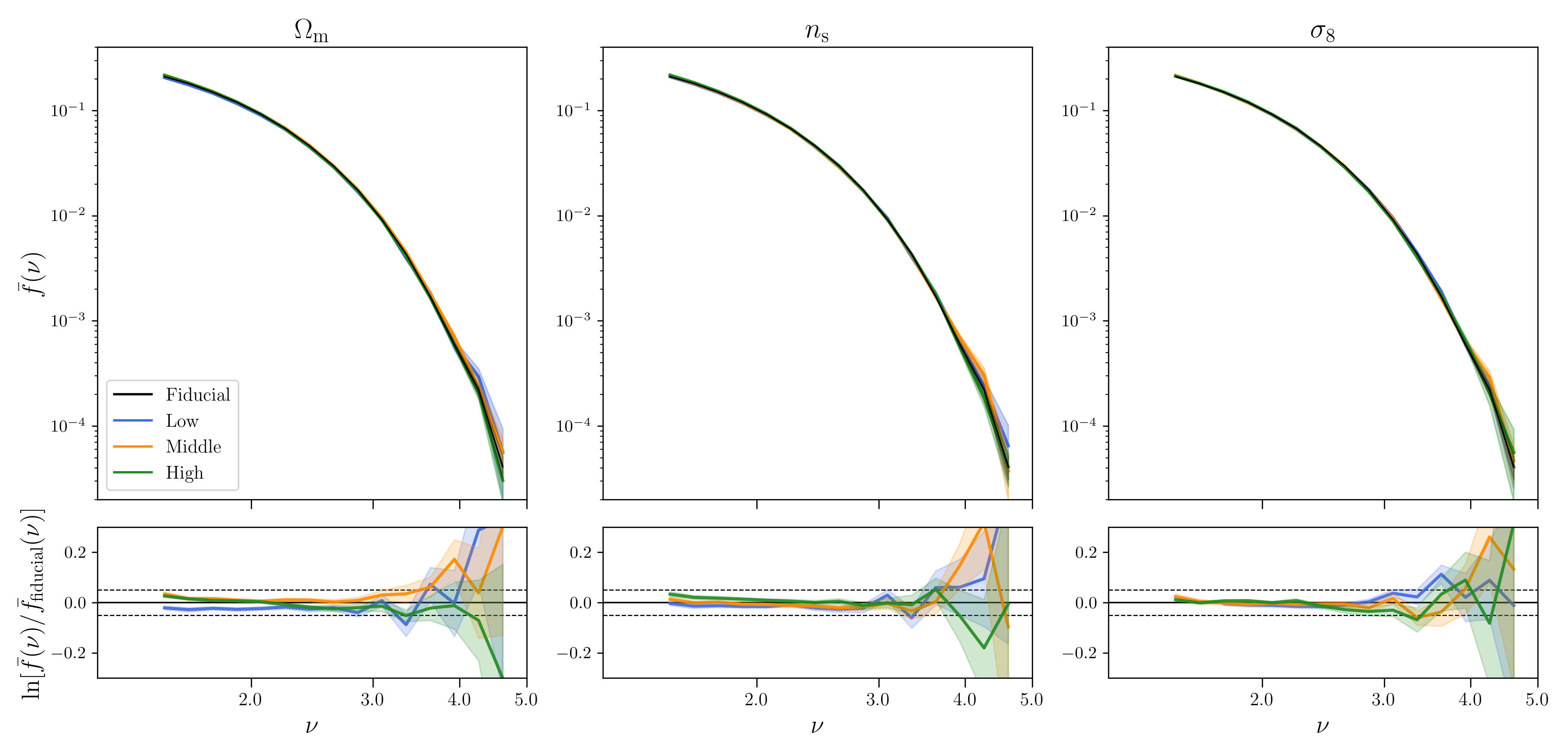}
    \caption{\textbf{Top:} Unweighted mean of the individual multiplicity functions interpolated onto a common peak-height grid of the Latin hypercube sample and Planck cosmology (referred to as Fiducial) from the Quijote suite. The shaded regions represent the standard error of the means. Each column divides the Latin hypercube boxes into three groups: lowest, middle, and highest values for the given cosmological parameters: $\Omega_{\rm m}$ (left), $n_{\rm s}$ (middle), and $\sigma_8$ (right).  \textbf{Bottom:} Natural-log ratios, $\ln[\bar f(\nu)/\bar f_{\rm fiducial}(\nu)]$. Shaded regions show approximate $\pm1\sigma$ uncertainties obtained by dividing the standard error of the group mean by $\bar f$ and adding the same uncertainty of the fiducial cosmology in quadrature. Dashed horizontal lines indicate approximately $\pm5\%$ deviations.}
    \label{fig:quijote_mf}
\end{figure*}

To perform this test, we sort the 100 Latin hypercube boxes into the 34 lowest, 33 middle, and 33 largest values for each of the three parameters we consider ($\Omegam, n_{\rm s}$, $\sigma_8$) and compute the $z=0$ multiplicity functions averaged over each of the three sets of boxes. We also average the multiplicity function of all 100 $z=0$ fiducial boxes. Figure~\ref{fig:quijote_mf} compares the resulting average multiplicity functions, with the bottom panels showing the ratio of the multiplicity functions relative to the fiducial boxes.  The widths of the bands in the plots correspond to the statistical uncertainties in the mean multiplicity functions and their ratios, respectively.  We detect non-universality of the halo mass function at high confidence --- e.g. the blue and green ratios for the ``$\Omega_{\rm m}$'' panel are clearly inconsistent with each other.  However, the mass functions are universal at the $\approx 5\%$ level (dashed horizontal lines in the ratio panels).  We caution, however, that differences in the dynamical mass of a halo are at the $\approx 2$--$4\%$ level as seen in the comparisons with Paper I and {\sc Sparta}, which could likely accommodate the level of non-universality evidenced in Figure~\ref{fig:quijote_mf}. A more detailed analysis of the mass function will be presented in future work.


%% file: sections/summary.tex
\section{Summary and Discussion}
\label{sec:discussion}

Following Paper I, we have defined a \it dynamical halo \rm as the collection of all particles orbiting their self-generated potential.  We have designed and developed a fast algorithm for generating dynamical halo catalogs from single-snapshot data by post-processing \rockstar\ halo catalogs. As an example, post-processing of the Quijote simulation boxes with $1024^3$ particles takes $\approx 10$ minutes on a single machine with 16 cores (32 threads).  \oasis\ relies on a radially dependent kinetic energy cut to separate orbiting from infalling particles. In this work we have shown that

\begin{itemize}
    \item The kinetic energy cut required to separate orbiting from infalling particles can be self-calibrated from the simulation data. 
    \item The linear cut in kinetic energy space is cosmology independent except for a logarithmic dependence of the slope of the cut on $\Omegam(z)$.
    \item The kinetic energy cut is mass independent, so long as radii are measured in units of $R_{\rm 200m}$, and velocities are measured in units of $v_{\rm 200m}$. 
    \item Comparisons with the {\sc Sparta} orbital classification from \citet{Vladimir2025} demonstrate the self-calibrated kinetic energy cut misses a small fraction of orbiting particles in the core, but that these particles can be readily included through a slight modification of the cut towards halo cores.
    \item The sharp cuts in kinetic energy can split substructures in two, with a fraction of the substructures labeled as orbiting, and a fraction labeled as infalling. Therefore, handling of substructures requires an additional step where we enforce that all particles within a substructure share the same classification.  We note that because substructures are classified as either parent or subhalos based on their orbital classification, \it the \oasis\ dynamical halo catalogs do not suffer from halo exclusion. \rm 
    \item Despite the fact that \oasis\ relies on single-snapshot data, the dynamical halo catalog generated with \oasis\ is in excellent agreement with the dynamical halo catalog we constructed in Paper I and the catalogs generated using {\sc Sparta}.
\end{itemize}

Having validated the \oasis\ algorithm, and as an illustration of the type of analyses enabled by our work, we applied \oasis\ to the Erebos and Quijote simulation suites to characterize the mass function of dynamical halos. Our principal results are as follows:
\begin{itemize}
    \item Contrary to our results in Paper I, we find that we cannot reasonably model the abundance of dark matter halos as a Press--Schechter function with a slowly varying threshold for collapse: such a model fails to describe the low-mass end of the halo mass function. If we restrict our analysis to the type of high-mass halos we analyzed in Paper I, we are able to reproduce those results.
    \item We demonstrated that a standard modification of the Press--Schechter form can describe the mass function of dynamical halos using a threshold for collapse of $\dsc = 1.54$ with 1.4\% intrinsic scatter, though the model becomes biased at high masses (see Figure~\ref{fig:Erebos_mf}). Despite the low value compared to the theory predictions, it is still in agreement with results from \citet{Shapiroetal99}.
    \item The mass function at $z=0$ is approximately universal with respect to cosmology at $\approx5\%$ accuracy, though we detect non-universality at high statistical significance (left panel, Figure~\ref{fig:quijote_mf}). The observed level of non-universality can be plausibly accommodated within the $\approx 2$--$4\%$ systematic error budget in the dynamical halo masses.
\end{itemize}

These results are illustrative of the type of analyses that are enabled by \oasis.  That is, we are now able to generate dynamical halo catalogs for large simulation suites, enabling us to characterize the cosmological dependence of the statistical properties of dynamical halos as a function of redshift and cosmology, a goal that we will turn to in our future work.
